\documentclass[reprint]{revtex4-2}
\usepackage[dvipsnames]{xcolor}
\usepackage{graphicx}
\usepackage{amsmath}
\usepackage{tablefootnote}
\usepackage{amssymb}
\usepackage{slashed}
\usepackage{hyperref}
\usepackage{enumerate}
\usepackage{tabularx}
\usepackage{bbold}

\usepackage{csvsimple}

\usepackage{lipsum}
\usepackage[normalem]{ulem}

\newcommand{\lagr}{\mathcal{L}}
\newcommand{\MeVfm}{{\text{ MeV/fm}^3}}
\newcommand{\MeV}{{\text{ MeV}}}
\newcommand{\GeV}{{\text{ GeV}}}
\newcommand{\fm}{{\text{ fm}^{-3}}}
\newcommand{\Mmax}{{M_\text{max}}}

\newcommand{\Mtrans}{{M_\text{trans}}}
\newcommand{\Msun}{{M_\odot}}
\newcommand{\Mchirp}{{\mathcal{M}_c}}
\newcommand{\NSmax}{{\text{NS}_\text{max}}}

\graphicspath{ {./Images/} }

\begin{document}

\title{Hybrid Stars with Different Amount of Quark Matter}

\author{Milena Albino}
\email{milena.albino@student.uc.pt}
\affiliation{Department of Physics, CFisUC, University of Coimbra, P-3004 - 516 Coimbra, Portugal}

\author{Tuhin Malik}
\email{tm@uc.pt}
\affiliation{Department of Physics, CFisUC, University of Coimbra, P-3004 - 516 Coimbra, Portugal}

\author{Márcio Ferreira}
\email{marcio.ferreira@uc.pt}
\affiliation{Department of Physics, CFisUC, University of Coimbra, P-3004 - 516 Coimbra, Portugal}

\author{Constança Providência}
\email{cp@uc.pt}
\affiliation{Department of Physics, CFisUC, University of Coimbra, P-3004 - 516 Coimbra, Portugal}

\date{\today}

\begin{abstract}
  In a previous work,
  we explored the possibility of neutron stars with large quark cores.
  Through Bayesian inference,
  we obtained large sets of equations of state
  imposing constraints regarding the properties of nuclear matter,
  mass-radius data from NICER,
  the GW170817 data from Virgo-LIGO collaboration, perturbative QCD and causality.
  We also imposed that the maximum mass neutron star ($\NSmax$) must have a large quark core.
  In this work, we obtain a new set of equations of state (SmallQ set)
  with the same constraints as before.
  However, we relaxed the last constraint,
  imposing only that $\NSmax$ must have quark core, regardless of its size.
  We compare the results with the set from our previous paper constrained by NICER and GW170817 observations NJL-GW.
  For both sets, we use the Relativistic Mean Field model
  with nonlinear meson interactions for the hadron phase
  and the Nambu--Jona-Lasinio model
  with four- and eight-quark interaction terms for the quark phase.
  The SmallQ set has higher values of phase transition density, with 3 -- 4.5 $\rho_0$,
  compared to the NJL-GW, with 2 -- 2.5 $\rho_0$.
  The new set describes NSs with smaller radius and smaller tidal deformability,
  and is more compatible with NICER and GW170817 data.
  A difference in the quark matter composition is observed,  with a larger quark strange fraction occurring in the SmallQ set.
  Within this set,  $2 \Msun$ NSs may contain little to no quark matter.
 It is shown that 
  there is  a non negligible fraction of EOS whose mass-radius curve presents a positive slope for 1.8 $\Msun$ stars, while for nucleonic stars this fraction is negligible and for hyperonic stars the slope is always negative.
\end{abstract}

\maketitle

\section{Introduction}
  Neutron stars (NSs) are the second most compact objects in the Universe.
  Their inner core can reach several densities above the nuclear saturation density
  ($\rho_0 = 0.16 \fm$),
  which cannot be reproduced by terrestrial experiments or
  described by first-principle calculations.
  Such extreme conditions may allow for the presence of exotic matter,
  as hyperons or quark matter.
  The latter case forms what is known as hybrid stars.
% In this work, we explore the latter case.

  The presence of exotic matter in NS's core
  may present different signatures, % in some properties,
  such as in the mass-radius diagram and in the tidal deformability.
  These quantities can be compared with the observational data
  of X-ray observations from NICER \cite{Riley:2019yda,miller_2019,miller_2021,Riley:2021pdl,salmi_2024,choudhury_2024,vinciguerra_2024} and the GW170817 data \cite{abbott_2018}
  generated by the Virgo-LIGO Collaboration.
  The observational data filter the EOS
  and may indicate whether a hybrid star is more likely than a hadron star
  \cite{passarella_2026, albino_2024, albino_2026}.
  Although the NS's radius data still remains of large uncertainty,
  %NICER data have a strong constraint on the mass, imposing that $\Mmax \gtrsim 2 \Msun$, to satisfy the  J0740+6620 observation \cite{salmi_2024,miller_2021}. 
  the J0740+6620 observation \cite{salmi_2024,miller_2021,Riley:2021pdl} already imposes a  strong constraint on the mass, $\Mmax \gtrsim 2 \Msun$. This constraint may be difficult to reconcile with GW170817
  since it requires a soft EOS while the $\Mmax$ constraint requires a stiff EOS.

  Other quantities may also indicate the presence of exotic matter,
  such as the profile of nuclear matter properties (NMP) \cite{albino_2026},
  polytropic index ($\gamma = d \ln P / d \ln \epsilon$)
  \cite{annala_2020},
  measure of conformability
  ($d_c = \sqrt{ \Delta^2 +\Delta'^2 }$,
  where $\Delta = 1/3 -P/\epsilon$ is the trace anomaly and
  $\Delta' = d \Delta / d \ln \epsilon$) \cite{annala_2023},
  and quantities related to binary NS mergers \cite{hammond_2026,guo_2025}.
  These methods have been used in agnostic models, whose  equations of state (EOS) are generated without assuming any microscopic behavior,
  and are constrained by NS data, ab-intio calculations at low (chEFT) and high density (pQCD). In \cite{annala_2020,annala_2023}, 
  %the authors used agnostic models and obtained evidence for the presence of quark matter in NSs.
   the conditions $\gamma < 1.75$ and $d_c < 0.2$ %,
%  where $\gamma = d \ln P / d \ln \epsilon$ is the polytropic index
%  and $d_c = \sqrt{ \Delta^2 +\Delta'^2 }$ is the measure of conformality,%
%  \footnote{
%    $\Delta = 1/3 -P/\epsilon$ is the trace anomaly and
%    $\Delta' = d \Delta / d \ln \epsilon$.
%  }
   were used to indicate quark matter behavior.%, respectively.

  Microscopic models generate the EOS from a Lagrangian,
  which contains information about the composition of matter.
  This is a clear advantage for probing quark matter in NSs compared with the agnostic models. In this work, we use the microscopic models approach to generate the EOS
  and compare it with the results from \cite{annala_2020,annala_2023}. We expect to understand which composition  may generate the different regions of phase space obtained in those works.

  It has been proposed that the composition of NS matter may also be inferred from the shape of the mass-radius or mass-tidal deformability relation.
  In fact, exotic matter may leave a strong imprint on the slope of the mass-radius diagram.
  In \cite{ferreira_2025}, the authors have  shown that NSs with only nucleonic  matter in their composition  predominantly
  present a negative slope of the mass-radius  curve at $M = 1.2, \; 1.4, \; 1.8 \Msun$,
  while NSs with hyperonic matter present a positive slope at small masses, $\lesssim 1.4 M_\odot$, and a negative slope for $M\gtrsim 1.8\, M_\odot$.
  In \cite{albino_2026}, we obtained a positive slope at $M = 1.2, \; 1.4, \; 1.6 \Msun$
  for $\gtrsim 90\%$ of the hybrid EOS. Even at $ 1.8 \Msun$ about 25\% of the stars still presented a positive slope.
  In addition to the radius, the  tidal deformability  derivatives may  also indicate the presence of exotic matter inside a NS \cite{bauswein_2026}.

 % metamodels vs microscopic models
%  There are two different approaches to probe the quark matter inside NSs:  through microscopic or agnostic models.  In the latter method, equations of state (EOS) are generated without assuming any microscopic behavior, and they are constrained by NS data, nuclear matter properties (NMP), and  perturbative QCD (pQCD) calculations.
  %
 %Although these conditions were chosen carefully,

  In a previous work \cite{albino_2026},
  we used Bayesian inference to obtain five different sets of equations of state (EOS): one set contains only hadron matter,
  and the other four sets contain hybrid EOS.
  The hadron phase is described by the Relativistic Mean Field (RMF) model
  with non-linear meson interactions \cite{malik_2023}
  and the quark phase is described by
  the Nambu--Jona-Lasinio (NJL) \cite{nambu_1961,klevansky_1992,hatsuda_1994}
  with added four- and eight-quark interaction terms
  or by the Mean Field Theory of QCD (MFTQCD) \cite{fogaca_2011}.
  Although the imposed constraints in the Bayesian inference
  had some differences between the sets,
  we have always imposed that the maximum mass NS generated by the EOS must have large quark cores.
  As a consequence, the phase transition occurs at low density: 2--2.5 $\rho_0$ for the three NJL sets,
  and 1--2 $\rho_0$ for the MFTQCD set. In addition, large quark cores require quite stiff nuclear EoS, much stiffer than predicted by nuclear EoS.

 In \cite{starcollaboration_2025}, the STAR Collaboration analyzes the results of the Au+Au reactions at a collision energy of $\sqrt{s_\text{NN}} = 3 \GeV$ in a dense medium.
  It is shown that  hadron interactions dominate in these conditions,
  indicating that hadron matter likely constitutes the medium.
  These results seem to indicate that these reactions can reach 3.6--4.0 $\rho_0$ and that the quark-hadron phase transition has to occur
  at higher densities according to \cite{yong_2022}.
  %In \cite{yong_2022}, the authors analyze the STAR Collaboration results and show that these reactions can reach 3.6--4.0 $\rho_0$. These may indicate that the quark-hadron phase transition occurs at higher densities.

  In this work, we obtained a new set of hybrid EOS (SmallQ set),
  with the hadron and quark phases described by the RMF and NJL models.
  We impose the same constraints from the NJL-GW set of our previous work \cite{albino_2026},
  but we relax the constraints on the phase transition density
  and on the difference between the central pressure of the $\NSmax$
  and the pressure at the phase transition density, which are both related to the quark core size.
  These resulted in NSs with small quark cores and
  a quark-hadron phase transition occurring at larger densities,
  which are more compatible with the results obtained in \cite{yong_2022}.
  % NS data
  Similar to the NJL-GW set from our previous work,
  the Bayesian inference also constrains the SmallQ set to satisfy
  nuclear matter properties (NMP),
  pQCD constraints at $7 \rho_0$ from \cite{gorda_2023,komoltsev_2023},
  GW170817 data from the Virgo-LIGO Collaboration \cite{abbott_2018}, and
  the pulsars
  J0030-0451 \cite{vinciguerra_2024,miller_2019},
  J0740-6630 \cite{salmi_2024,miller_2021}, and
  J0437-4715 \cite{choudhury_2024,reardon_2024} from NICER.
The pQCD constraint is imposed at $7\rho_0$ because, for the large majority of our EOS, the central density of the maximum-mass NS lies below this value (the 90\% CI upper limit of $\rho_\text{max}$ is $\approx 6\rho_0$, see Table~\ref{tab:NS_prop}), so that the whole density range realised in stable NSs is required to be consistent with pQCD.

  % signatures of quark matter
  By comparing the SmallQ (small quark core) set with
  the NJL-GW (large quark core) and
  RMF (no quark core, only hadron matter)
  sets from our previous paper,
  we aim to analyze the implications of the presence of quark matter inside NSs and identify possible  properties capable of  distinguishing NSs with different quark core sizes.

  This article is organized as follows:
  Sec. \ref{sec:eos} summarizes the theoretical models for the hadron and quark phases.
  Sec. \ref{sec:bayesian} describes the Bayesian inference,
  including the prior and likelihood distributions used.
  In Sec. \ref{sec:results}, we discuss the results of the EOS,
  the properties of NSs,
  the influence on the quark-hadron phase transition, and nuclear matter properties. A comparison with NS with large quark cores is discussed, as well as possible signatures of the presence of quark core.
  In Sec. \ref{sec:conclusion}, we summarize the conclusions.

\section{Equations of State}
\label{sec:eos}
    In this section,
    we describe the models used to obtain the equations of state (EOS) of the hybrid stars.
    For the hadron phase,
    we adopted the Relativistic Mean Field (RMF) model,
    and for the quark phase,
    we used the Nambu--Jona-Lasinio (NJL).
    The quark-hadron phase transition is build by using the Maxwell construction.

    \subsection{Hadron phase}
        In this work, the hadron phase is described by the
        Relativistic Mean Field (RMF) model with nonlinear meson interaction terms.
        In this model, nuclear interactions are mediated by the
        exchange of three mesons:
        the scalar-isoscalar, vector-isoscalar and vector-isovector mesons
        ($\sigma$, $\omega$ and $\varrho$, respectively).
        The Lagrangian is given by
        \begin{equation}
            \mathcal{L} = \mathcal{L}_N +\mathcal{L}_M +\mathcal{L}_{NL},
        \end{equation}
        where
        \begin{align}
            \mathcal{L}_N &= \bar{\Psi}
                \left[
                    \gamma^\mu \left(
                        i \partial_\mu -g_\omega \omega_\mu
                        -g_\rho \boldsymbol{t} \cdot \boldsymbol{\rho}_\mu
                    \right)
                    \right. \nonumber \\
                    &\quad \left.
                    -(m -g_\sigma \phi)
                \right] \Psi,  \\
            \mathcal{L}_M &= \frac{1}{2}
                \left[
                    \partial_\mu \sigma \partial^\mu \sigma
                    -m_\sigma^2 \sigma^2
                \right] \nonumber \\
                &\quad
                -\frac{1}{4} \omega_{\mu\nu} \omega^{\mu\nu}
                +\frac{1}{2} m_\omega^2 \omega_\mu \omega^\mu \nonumber \\
                &\quad
                -\frac{1}{4} \boldsymbol{\varrho}_{\mu\nu} \cdot \boldsymbol{\varrho}^{\mu\nu}
                +\frac{1}{2} m_\varrho^2 \boldsymbol{\varrho}_\mu \cdot \boldsymbol{\varrho}^\mu, \\
            \mathcal{L}_{NL} &=
                -\frac{1}{3} b m g_\sigma^3 (\sigma)^3
                -\frac{1}{4} c g_\sigma^4 (\sigma)^4
                +\frac{\xi}{4!} g_\omega^4 (\omega_\mu \omega^\mu)^2 \nonumber \\
                &\quad
                +\Lambda_\omega g_\varrho^2 \boldsymbol{\varrho}_\mu \cdot \boldsymbol{\varrho}^\mu
                g_\omega^2 \omega_\mu \omega^\mu,
        \end{align}
        where the Dirac spinor $\Psi$, which has bare mass $m$,
        describes the proton and neutron fields.
        The Dirac matrices are represented by the $\gamma^\mu$,
        and $\bm{t} = \bm{\tau}/2$ is the isospin operator,
        where $\tau$ are the Pauli matrices.
        $g_i$ and $m_i$ are the coupling constants and masses of the mesons
        $i = \sigma, \omega, \varrho$
        and $\omega_{\mu\nu} = \partial_\mu \omega_\nu -\partial_\nu \omega_\mu$
        (similar for $\varrho$).

        By calculating the equations of motion and
        applying the relativistic mean field approximation \cite{glendenning_1996},
        the meson fields are expressed as
		\begin{align}
			\sigma = &\frac{g_\sigma}{m_{\sigma,\text{eff}}^2} \sum_i \rho_i^S,
			\\
			&m_{\sigma,\text{eff}}^2 =
            m_\sigma^2 +b m g_\sigma^3 \sigma +c g_\sigma^4 \sigma^2, \\
			\omega = &\frac{g_\omega}{m_{\omega,\text{eff}}^2} \sum_i \rho_i,
			\\
			&m_{\omega,\text{eff}}^2 = m_\omega^2
	    		+\frac{\xi}{3!} g_\omega^4 \omega^2
	            +2 \Lambda_\omega g_\varrho^2 g_\omega^2 \varrho^2, \\
			\varrho = &\frac{g_\varrho}{m_{\varrho,\text{eff}}^2} \sum_i t_3 \rho_i,
			\\
			&m_{\varrho,\text{eff}}^2 = m_\varrho^2
			      +2 \Lambda_\omega g_\varrho^2 g_\omega^2 \omega^2,
		\end{align}
	    where $t_3 = \pm 1/2$ for protons and neutrons, respectively;
        and $\rho^S$ and $\rho$ are the scalar and number densities. 
%        Calculating the equation of motion of the nucleon doublet $\Psi$,
%        we obtain the baryon chemical potential
%        \begin{equation}
%            \mu_B = E(\vec{k}) +g_\omega \omega +g_\varrho t_3 \varrho,
%        \end{equation}
%        where
%        \begin{equation}
%            E(\vec{k}) = \sqrt{\vec{k}^2 +(m -g_\sigma \sigma)^2}.
%        \end{equation}
        To apply this model in NSs condition,
        we impose chemical equilibrium and charge neutrality conditions,
        given by
        \begin{align}
          \mu_p &= \mu_n -\mu_e, \\
          \mu_\mu &= \mu_e, \\
          0 &= \sum_{i = p,e,\mu} q_i \rho_i,
        \end{align}
        where $q_i$ is the electric charge.
        The electron and muon are included as a free Fermi gas.
        The parameters
        $g_\sigma$, $g_\omega$, $g_\varrho$,
        $b$, $c$, $\xi$ and $\Lambda_\omega$
        are sampled in Bayesian inference.

    \subsection{Quark phase}

     To describe the quark phase in NSs,
      we use the 3-flavor Nambu--Jona-Lasinio (NJL) model
      with the 't-Hooft term and four- and eight-quark interaction terms.
      The NJL model reproduces all the global symmetries of QCD
      and the sponraneous chiral symmetry breaking manifestations.
      It is a suitable model for low energy regimes, as in the NSs conditions.
      The Lagrangian is given by
      \begin{align}
        \mathcal{L} &=
          \bar{\psi} (i \gamma_\mu \partial^\mu -m +\mu \gamma^0) \psi  \nonumber \\
          &\quad +\frac{G}{2}
            \left[
              (\bar{\psi} \lambda_a \psi)^2
              +(\bar{\psi} i \gamma^5 \lambda_a \psi)^2
            \right] \nonumber \\
          &\quad +\kappa
            \left\{
              \det_f[\bar{\psi} (1 +\gamma^5) \psi]
              +\det_f[\bar{\psi} (1 -\gamma^5) \psi]
            \right\} \nonumber \\
          &\quad +\mathcal{L}_\text{int},
          \label{eq:NJL_L}
      \end{align}
      where $m$ and $\mu$ are the quark current masses and chemical potential matrices,
      $\lambda_a$ (with $a = 1, 2, \dots, 8$ are the Gell-Mann matrices,
      $\lambda_0 = \sqrt{2/3} \mathbb{1}$,
      and the interaction terms are given by
     \begin{align}
        \mathcal{L}_\text{int}
        &= -G_\omega
          \left[
            (\bar{\psi} \gamma^\mu \lambda_0 \psi)^2
            +(\bar{\psi} \gamma^\mu \gamma^5 \lambda_0 \psi)^2
          \right] \nonumber \\
        &\quad -G_\rho
          \sum_{a=1}^8
          \left[
            (\bar{\psi} \gamma^\mu \lambda_a \psi)^2
            +(\bar{\psi} \gamma^\mu \gamma^5 \lambda_a \psi)^2
          \right] \nonumber \\
        &\quad -G_{\omega\omega}
          \left[
            (\bar{\psi} \gamma^\mu \lambda_0 \psi)^2
            +(\bar{\psi} \gamma^\mu \gamma^5 \lambda_0 \psi)^2
          \right]^2 \nonumber \\
        &\quad -G_{\sigma\omega}
          \sum_{a=0}^8
          \left[
            (\bar{\psi} \lambda_a \psi)^2
            +(\bar{\psi} i \gamma^\mu \gamma^5 \lambda_a \psi)^2
          \right]
          \nonumber \\ &\quad \times
          \left[
            (\bar{\psi} \gamma^\mu \lambda_0 \psi)^2
            +(\bar{\psi} \gamma^\mu \gamma^5 \lambda_0 \psi)^2
          \right] \nonumber \\
        &\quad -G_{\rho\omega}
          \sum_{a=1}^8
          \left[
            (\bar{\psi} \gamma^\mu \lambda_a \psi)^2
            +(\bar{\psi} \gamma^\mu \gamma^5 \lambda_a \psi)^2
          \right]
          \nonumber \\ &\quad \times
          \left[
            (\bar{\psi} \gamma^\mu \lambda_0 \psi)^2
            +(\bar{\psi} \gamma^\mu \gamma^5 \lambda_0 \psi)^2
          \right].
        \label{eq:NJL_int}
      \end{align}
      These interaction terms were studied in \cite{}
      and have an important role in describing the observational data of NSs.

      By linearizing the Lagrangian and applying the mean field approximation,
      we obtain the gap equations
      \begin{align}
        \tilde{m}_i &= m_i -2 G \sigma_i -2 \kappa \sigma_j \sigma_k \nonumber \\
          &\quad
          +\frac{8}{3} G_{\sigma\omega} (\rho_i^2 +\rho_j^2 +\rho_k^2) \sigma_i \\
        \tilde{\mu}_i &= \mu_i -
          \frac{4}{3} G_\omega (\rho_i +\rho_j +\rho_k)
          -\frac{4}{3} G_\rho (2 \rho_i -\rho_j -\rho_k) \nonumber \\
        &\quad
          -\frac{16}{9} G_{\omega\omega} (\rho_i +\rho_j +\rho_k)^3 \nonumber \\
        &\quad
          -\frac{8}{3} G_{\sigma\omega} (\sigma_i^2 +\sigma_j^2 +\sigma_k^2)
            (\rho_i +\rho_j +\rho_k) \nonumber \\
        &\quad
          -\frac{8}{9} G_{\rho\omega} (\rho_i +\rho_j +\rho_k) \nonumber \\
        &\quad
          \times (4 \rho_i^2 +\rho_j^2 +\rho_k^2
          -\rho_i \rho_j -\rho_i \rho_k -4 \rho_j \rho_k),
      \end{align}
      with $i \neq j \neq k \in \{u, d, s\}$,
      $\tilde{m}$ and $\tilde{\mu}$ are the effective mass and
      effective chemical potential, respectively,
      and the quark condensates are defined as
      $\sigma_i = \left< \bar{\psi}_i \psi_i \right>$
      and
      $\rho_i = \left< \psi^\dag_i \psi_i \right>$.
      We also obtain the grand canonical potential,
      in which, at zero temperature, is given by
      \begin{align}
        \Omega
          &= \Omega_0
            +G
            \left(
              \sigma_u^2 + \sigma_d^2 + \sigma_s^2
            \right)
            +4 \kappa \sigma_u \sigma_d \sigma_s \nonumber \\
          &\quad
            -\frac{2}{3} G_\omega (\rho_u + \rho_d + \rho_s)^2 \nonumber \\
          &\quad
            -\frac{4}{3} G_\rho
            (\rho_u^2 +\rho_d^2 +\rho_s^2 -\rho_u \rho_d -\rho_u \rho_s -\rho_d \rho_s)
          \nonumber \\ &\quad
            -\frac{4}{3} G_{\omega\omega} (\rho_u + \rho_d + \rho_s)^4 \nonumber \\
          &\quad
            -4 G_{\sigma\omega}
            \left( \sigma_u^2 + \sigma_d^2 + \sigma_s^2 \right)
            \left( \rho_u + \rho_d + \rho_s \right)^2
          \nonumber \\
          &\quad -\frac{8}{3} G_{\rho\omega}
            \left( \rho_u + \rho_d + \rho_s \right)^2 \nonumber \\
          &\qquad \times
            \left(
              \rho_u^2 + \rho_d^2 + \rho_s^2
              -\rho_u \rho_d -\rho_u \rho_s -\rho_d \rho_s
            \right) \nonumber \\
          &\quad -\frac{3}{\pi^2}
            \sum_{f=u,d,s} \int_{k_{F_f}}^\Lambda dp p^2 E_f
            -\frac{1}{\pi^2} \sum_{f=u,d,s} \tilde{\mu}_f k^3_{F_f},
      \end{align}
      where $E_f = \sqrt{p^2 +\tilde{m}_f^2}$ is the energy.
      The $\Omega_0$ is a constant set to vanish the grand canonical potential in the vacuum,
      and
      $k_{F,f} = \sqrt{\tilde{\mu}_f^2 - \tilde{m}_f^2}$
      is the Fermi momentum.
      Since the NJL is a non-renormalizable model,
      we use the 3-momentum cutoff scheme ($\Lambda$).
      By imposing that
      $d \Omega / d\sigma_i = d \Omega / d\rho_i = 0$,
      we obtain the expressions for the quark condensates at $T = 0$:
      \begin{align}
        \sigma_f &=
          -\frac{3}{\pi^2}
          \int_{k_{F_f}}^\Lambda dp p^2 \frac{\tilde{m}_f}{E_f}, \\
        \rho_f &= \frac{1}{\pi^2} k_{F_f}^3.
      \end{align}

      From the grand canonical potential, we obtain the EoS of the NJL model.
      To satisfy the NSs conditions,
      we impose the chemical potential equilibrium and charge neutrality condition,
      given by
      \begin{align*}
        \mu_d & = \mu_u +\mu_e, \\
        \mu_s &= \mu_d, \\
        0 &= \sum_{i=u,d,s,e} q_i \rho_i,
      \end{align*}
      where $q_i$ is the electric charge.
      The electron contribution is added as a free Fermi gas.
      Also, since the NJL is normalized in the vacuum,
      we set $P \to P +B$, where $B$ is a constant
      and has a strong influence of the phase transition location.

      The parameters $G$, $\kappa$, $m_i$ and $\Lambda$
      are set to satisfy the mass and decay constant data from
      $\pi^\pm$, $K^\pm$, $\eta$ and $\eta'$ in Table \ref{tab:fixed_njl}.
      The parameters $\xi_\omega$, $\xi_\rho$, $\xi_{\omega\omega}$,
      $\xi_{\sigma\omega}$, $\xi_{\rho\omega}$, defined as
      \begin{align}
        \xi_i &= 2 \frac{G_i}{G}, \quad i = \omega, \rho \\
        \xi_{jk} &= 16 \frac{G_{jk}}{G^4}, \quad jk = \omega \omega, \sigma \omega, \rho \omega,
      \end{align}
      and $B$
      are sampled through Bayesian analysis.
      \begin{table}[ht]
        \centering
        \setlength{\tabcolsep}{7.pt}
        \renewcommand{\arraystretch}{1.2}
        \begin{tabular}{ccccc}
          \hline \hline
          $\Lambda$ (MeV) & $m_{u,d}$ (MeV) & $m_s$ (MeV) & $G \Lambda^2$ & $\kappa \Lambda^5$ \\
          \hline
          623.58 & 5.70 & 136.60 & 3.34 & -13.67 \\
          \hline
        \end{tabular}
        \caption{Fixed parameters for NJL model,
          set to satisfy the experimental data from \cite{olive_2014}.}
        \label{tab:fixed_njl}
      \end{table}

\section{Bayesian Inference}
\label{sec:bayesian}

\begin{table*}[ht]
    \centering
    \setlength{\tabcolsep}{12.pt}
    \renewcommand{\arraystretch}{1.2}
    \begin{tabular}{ccccccc}
        \hline \hline
%        \multicolumn{7}{c}{Set NJL-GW}                                              \\
        \multicolumn{3}{c}{NJL}             &  & \multicolumn{3}{c}{RMF}            \\ 
        \hline 
        Parameters              & min & max &  & Parameters       & min    & max    \\
        \hline
        $\xi_\omega$            & 0   & 0.5 &  & $g_\sigma$       &  8     & 11     \\
        $\xi_\rho$              & 0   & 1   &  & $g_\omega$       & 10     & 14     \\
        $\xi_{\omega\omega}$    & 0   & 30  &  & $g_\rho$         &  9.546 & 15     \\
        $\xi_{\sigma\omega}$    & 0   & 8   &  & BB               &  1.500 &  3.500 \\
        $\xi_{\rho\omega}$      & 0   & 50  &  & CC               & -4.627 & -1.500 \\
        $B$ (MeV/fm$^3$)        & 0   & 30  &  & $\xi$            &  0     & 0.016 \\
                                &     &     &  & $\Lambda_\omega$ &  0     & 0.103 \\
        \hline
    \end{tabular}
    \caption{The lowest and highest values for the uniform distribution prior
    used for the NJL-GW and SmallQ sets.
    We defined $BB = b \times 10^3$ and $CC = c \times 10^3$.}
    \label{tab:prior}
\end{table*}

Bayesian inference is a powerful statistical tool
that estimates the parameters considering prior information and data constraints.
By setting the values of the model parameters ($\Theta$),
Bayesian inference applies the Bayes' theorem, given by
\begin{equation}
  P(\Theta | D) = \frac{\lagr(D | \Theta) \pi(\Theta)}{Z},
\end{equation}
where $D$ is the data used as a constraint.
On the right side of the equation,
we have the likelihood distribution ($\lagr(D | \Theta)$),
which is the probability that the parameters $\Theta$ satisfies the data,
and the prior distribution ($\pi(\Theta)$),
that is the probability of the parameters $\Theta$ are true
before considering any information of the data.
We also have the evidence ($Z$) as a normalization factor.
On the left side of the equation,
we have the posterior distribution,
which gives the combination of the probability
of the prior and likelihood probabilities.
This gives us the quality of the parameters $\Theta$
in accordance with the imposed constraints.
The values of $\Theta$ are sampled and
a large set of parameters with high posterior probability is returned.
Different techniques of sampling can be applied.
In this work, we used the Bayesian inference library BILBY \cite{ashton_2019}
with the pyMultiNest package \cite{buchner_2014,buchner_2023}, which is
based on the nested sampling method.

To apply the Bayesian inference,
in addition to the sampling method,
we need to define the prior and likelihood functions.
For the prior, we used a uniform distribution
with the minimum and maximum values defined by Table \ref{tab:prior}.
For the likelihood,
we consider the nuclear matter properties (NMP)
and pure neutron matter (PNM) data (see Table \ref{tab:NMP}),
the x-ray NICER data from
J0030-0451 \cite{vinciguerra_2024,miller_2019},
J0740-6630 \cite{salmi_2024,miller_2021},
J0437-4715 \cite{choudhury_2024,reardon_2024}, and
the gravitational wave (GW) data from
GW170817 \cite{abbott_2018}, and
the pQCD constraint from \cite{gorda_2023,komoltsev_2023} at 7 $\rho_0$.
We also imposed some constraints to ensure a phase transition
(see \cite{albino_2026} for more details).
The NMP are the coefficients of the expansion of the energy per nucleon around the saturation density $\rho_0$,
\begin{align}
&EA(\rho,\delta) \simeq EA_\text{SNM}(\rho) +S(\rho)\, \delta^2,
\label{eq:nmp_expansion} \\
&EA_\text{SNM}(\rho) = EA_0 +\frac{K_0}{2} x^2 +\frac{Q_0}{6} x^3, \\
&S(\rho) = J_{\text{sym},0} +L_{\text{sym},0}\, x +\frac{K_{\text{sym},0}}{2} x^2 +\frac{Q_{\text{sym},0}}{6} x^3,
\end{align}
where $x = (\rho -\rho_0)/(3\rho_0)$ and $\delta = (\rho_n -\rho_p)/\rho$ is the isospin asymmetry.
$EA_0$ is the binding energy per nucleon, $K_0$ the incompressibility and $Q_0$ the skewness of symmetric nuclear matter,
while $J_{\text{sym},0}$, $L_{\text{sym},0}$, $K_{\text{sym},0}$ and $Q_{\text{sym},0}$ are the symmetry energy and its slope, curvature and skewness at saturation.
Saturation is imposed through $dEA/d\rho|_{\rho_0} = 0$, and $EA_\text{PNM}$ is the energy per neutron of pure neutron matter.
Only the quantities listed in Table~\ref{tab:NMP} enter the likelihood; the remaining coefficients ($Q_0$, $L_{\text{sym},0}$, $K_{\text{sym},0}$ and $Q_{\text{sym},0}$) are predictions of the model.

\begin{table*}[ht]
    \centering
    \setlength{\tabcolsep}{12.pt}
    \renewcommand{\arraystretch}{1.2}
    \begin{tabular}{ccccccc}
        \hline \hline 
        \multicolumn{3}{c}{NMP}              &  & \multicolumn{3}{c}{PNM}                    \\ 
        Quantity & value/band & Ref.         & & Quantity & value/band & Refs. \\
        \hline
        $\frac{dEA}{d\rho}$ & 0              &
        & &
        $EA_{PNM}(\rho = 0.05)$ & $6.8 \pm 1.02$ & \cite{Huth:2021bsp} \\
        $EA_0$              & $-16 \pm 0.02$ & \cite{PhysRevC.90.055203} 
        & & 
        $EA_{PNM}(\rho = 0.10)$ & $10.5 \pm 1.97$ & \cite{Huth:2021bsp} \\
        $K_0$               & $230 \pm 30$   & \cite{PhysRevLett.95.122501,Shlomo:2006ole}
        & &
        $EA_{PNM}(\rho = 0.15)$ & $15.3 \pm 3.44$ & \cite{Huth:2021bsp} \\
        $J_{sym,0}$         & $32.5 \pm 1.8$ & \cite{PhysRevC.104.065804}
        & &
        & \\
        \hline
    \end{tabular}
    \caption{Constraints on the nuclear matter properties (NMP) and pure neutron matter (PNM) imposed in the likelihood.
NMP (defined in Eq.~(\ref{eq:nmp_expansion})): saturation condition $dEA/d\rho|_{\rho_0} = 0$,
binding energy per nucleon $EA_0$, incompressibility $K_0$ and symmetry energy $J_{\text{sym},0}$ at saturation.
PNM: energy per neutron $EA_\text{PNM}$ at $\rho = 0.05$, $0.10$ and $0.15$~fm$^{-3}$ from chiral effective field theory.
All energies are in MeV.}
    \label{tab:NMP}
\end{table*}

In \cite{albino_2026}, we obtained five sets of EoS that reproduce NS with large quark core.
In this work, we compare the NJL-GW from \cite{albino_2026}
with the new set obtained in this paper, the SmallQ set,
that reproduces $\NSmax$ with small quark core.
The differences between these two sets are the
constraints related to the size of the quark cores.
For the SmallQ set,
we did not apply the range of phase transition constraint ($\mathcal{L}_\text{phT}$).
We also relaxed the quark inside $\Mmax$ constraint ($\mathcal{L}_\text{Qmax}$).
This likelihood establishes a minimum value for the difference between the
pressure at $\NSmax$ ($P_\text{max}$) and $P_\text{trans}$.
For the NJL-GW set, we defined this minimum value as
$P_\text{max}-P_\text{trans} = 100 \MeVfm$,
while for the SmallQ,
we set $P_\text{max}-P_\text{trans} = 5 \MeVfm$.
In this way, Bayesian inference still favors $\NSmax$ with a quark core,
but with no restriction of its size.

\section{Results}
\label{sec:results}
In this section, we first discuss the  EOS properties, followed by a discussion of several NS properties  and hadron-quark phase transition properties. We finish by discussing the implications of the existence of a quark core on the properties of  nuclear matter.

    \subsection{Results of the EoS}

\begin{figure}[ht]
        \centering
        \includegraphics[width=.9\linewidth]{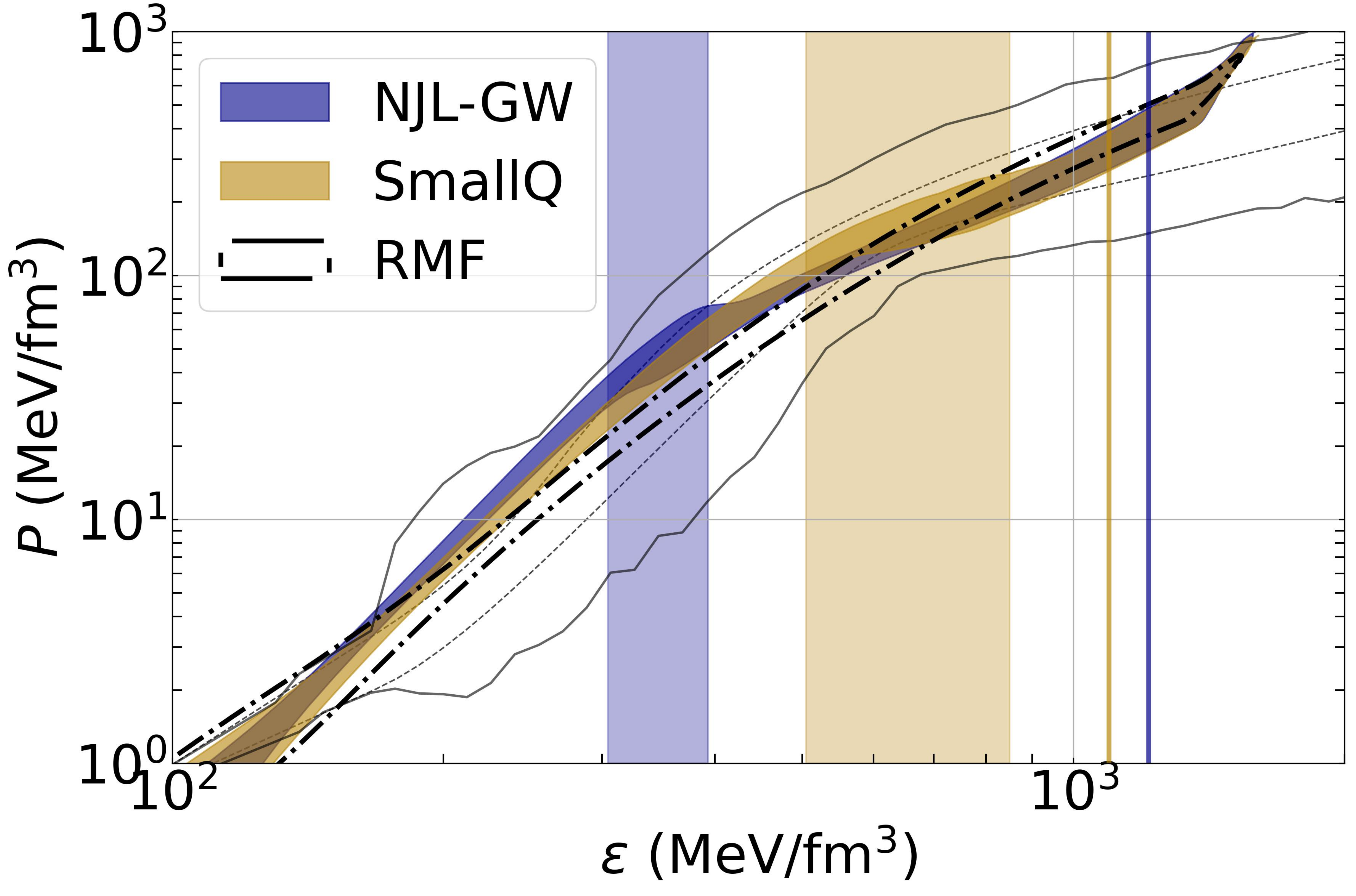}
        \caption{Pressure as a function of energy density
            of the 90\% CIs of the sets NJL-GW (purple), SmallQ (yellow) and RMF (black dash-dot).
            The vertical lines mark the 90\% CI maximum of the
            central energy density in the maximum mass NS ($\epsilon_\text{max}$).
            The vertical bands represent the 90\% CI interval for
            the energy density at the phase transition ($\epsilon_{H,\text{trans}}$).}
        \label{fig:PxE}
    \end{figure}

    We start by comparing the EOS distributions obtained for SmallQ with the ones obtained for NJL-GW. 
	% Figure P x \epsilon
	Figure \ref{fig:PxE} shows the 90\% CIs of pressure as a function of the energy density for the
	NJL-GW (purple), SmallQ (yellow), and RMF (black dash-dot) sets.
	The vertical lines mark the 90\% CI maximum value of the central energy density of the maximum mass NS.
	The vertical bands represent the 90\% CI of the hadron phase transition point ($\epsilon_\text{H,trans}$).
    
	Relaxing the constraint on the size of the quark core
	significantly increases the density localization of the phase transition point, shifting it to larger densities. This indicates that the data used to constrain the models prefer a smaller quark core. In addition, with respect to  the NJL-GW Set, the central energy density decreases by about $\sim 100 \MeVfm$, and the pressure decreases in the region  $\epsilon\lesssim 400 \MeVfm$.
    However, due to the phase transition to quark matter at high energy densities,
    the pressure of the SmallQ Set remains greater than the pressure of the NJL-GW Set for $\gtrsim 400 \MeVfm$, while the phase transition does not occur.
    
	We also compare our results with the model-independent results from \cite{Annala:2021gom,altiparmak_2022},
	which used a model independent speed of sound parametrization for the EOS.
	This is represented in Figure \ref{fig:PxE} by the gray solid (full) and dotted (90\% CI) contours.
	Although both sets lie inside the full model-independent envelope,
	the SmallQ set is more compatible with the 90\% CI contour than the NJL-GW set. 

	% Figure c_s^2 x \rho_B
    \begin{figure}[ht]
        \centering
        \includegraphics[width=.9\linewidth]{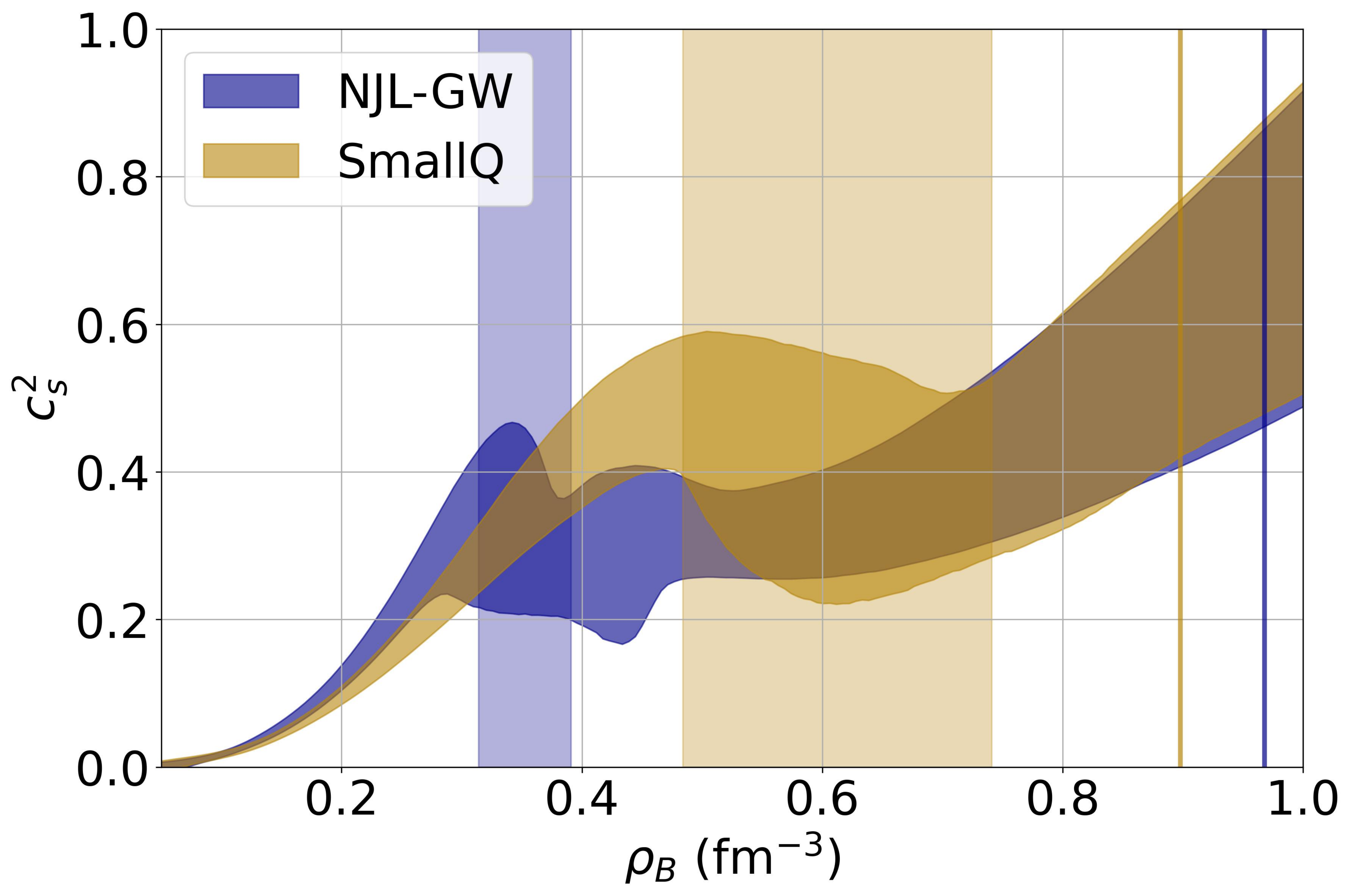}
        \caption{Speed of sound as a function of baryon density:
           the 90\% CIs for the sets NJL-GW (purple) and SmallQ (yellow).
            The vertical lines mark the 90\% CI maximum of the
            central energy density in the maximum mass NS ($\epsilon_\text{max}$).
            The vertical bands represent the 90\% CI interval for
            the energy density at the phase transition ($\epsilon_{H,\text{trans}}$).}
        \label{fig:cs2}
    \end{figure}

	In Figure \ref{fig:cs2}, we show  the 90\% CIs of the squared speed of sound versus the baryon density.
	Below $0.2 \fm$ and above $0.7 \fm$, the two sets have the same behavior, with the NJL-GW set having slightly higher $c_s^2$ values in the former region. However, between the $0.2$ and $0.7 \fm$, the sets produce different results.
	The NJL-GW curve has two bumps:
	the first one occurs due to the phase transition
	and the second occurs due to the onset of the quark strange.
	The SmallQ curve has only one bump
	because the phase transition occurs after the onset of the quark strange.
    Due to this difference in the phase transition point,
    the SmallQ Set results in larger values for the speed of sound
    in the intermediate region.
    % Figure dc, gamma, trace anomaly
    \begin{figure}[ht]
		\centering
		\includegraphics[width=0.9\linewidth]{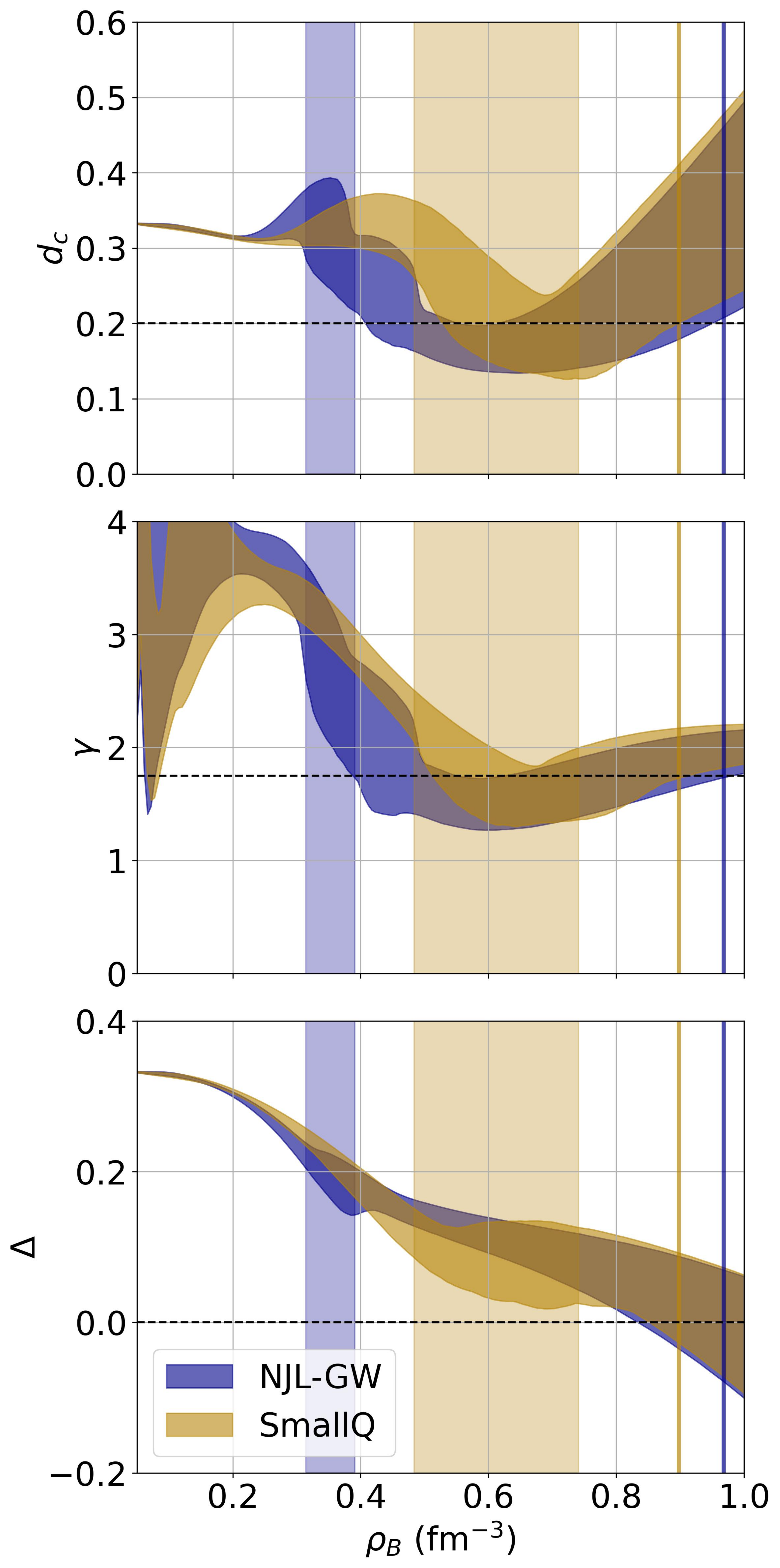}
		\caption{Measure of conformability ($d_c$),
		      polytropic index ($\gamma$) and 
		      trace anomaly ($\Delta$) as a function of density.}
		\label{fig:trace}
    \end{figure}

	Figure \ref{fig:trace} shows three quantities  related to the EOS as functions of density, which have been introduced in \cite{annala_2020,Fujimoto:2022ohj,annala_2023} to discuss  quark deconfinement.
	From bottom to top:
	the normalized trace anomaly, defined as $\Delta = 1/3 -P/\epsilon$ \cite{Fujimoto:2022ohj};
	the polytropic index $\gamma = d \ln P / d \ln \epsilon$ \cite{annala_2020}; and
	the measure of conformability $d_c = \sqrt{\Delta^2 +\Delta'^2}$,
	where $\Delta' = d \Delta / d \ln \epsilon$ \cite{annala_2023}. 
	%In \cite{annala_2020,annala_2023}, the authors study matter deconfinement through model-independent EoS.
	The regions defined by $d_c < 0.2$  \cite{annala_2023} or with $\gamma < 1.75$ \cite{annala_2020} were proposed as identifying the presence of  quark  deconfined matter.
    
	In the top plot, $d_c$  exhibits  a decreasing behavior for both sets
	within their respective phase transition regions and enters the region $d_c<0.2$ as quark matter. 
	However, %$d_c$ remains larger than 0.2 even in quark matter.
	for  $\rho_B \gtrsim 0.7 \fm$, the $d_c$ exhibits increasing behavior, returning to the region of $d_c > 0.2$ due to the presence of a strong repulsive vector potential in the NJL model.
	In the middle plot, the $\gamma$ functions decrease
	for $\rho_B \lesssim 0.5$ for both sets, and at $\gamma < 1.75$, they are both in a quark deconfined phase. However, 
	at $\rho_B \gtrsim 0.5$, the values of $\gamma$ increase above 1.75
	and remain constant at $\approx 2$.
    For the $d_c$ and $\gamma$ results,
    the SmallQ set has more compatible results  with the definitions in \cite{annala_2020,annala_2023},
    since the phase transition to quark deconfined matter occurs at $d_c \sim 0.2$ and $\gamma \sim 1.75$.
    The trace anomaly is shown in the bottom plot.
    For both sets, the majority of the values of $\Delta$ are positive.

  \begin{table*}[ht]
    \centering
    \setlength{\tabcolsep}{12.pt}
    \renewcommand{\arraystretch}{1.2}
    \begin{tabular}{ccccccc}
      \hline \hline
      & \multicolumn{3}{c}{Set SmallQ} & \multicolumn{3}{c}{Set NJL-GW} \\
      quant & median & min & max & median & min & max \\
      \hline
      $\rho_\text{trans}$            & 0.592   & 0.484   & 0.741    & 0.362    & 0.314   & 0.391    \\
      $P_\text{trans}$               & 182.666 & 121.281 & 276.928  & 55.989   & 33.814  & 77.024   \\
      $\epsilon_\text{H,trans}$      & 644.373 & 505.151 & 850.219  & 358.957  & 304.433 & 393.36   \\
      $\epsilon_\text{Q,trans}$      & 806.523 & 630.096 & 977.917  & 395.183  & 334.473 & 458.06   \\
      $\Delta \epsilon_\text{trans}$ & 151.216 & 92.699  & 195.299  & 36.207   & 19.691  & 76.780   \\
      $\rho_s$                       & 0.350   & 0.295   & 0.460    & 0.455    & 0.380   & 0.500    \\
      $Y_{s,\text{max}}$             & 0.2487  & 0.1909  & 0.2795   & 0.2218   & 0.1839  & 0.2542   \\
      $M_\text{max}$                 & 2.168   & 2.047   & 2.282    & 2.108    & 1.993   & 2.212    \\
      $R_\text{max}$                 & 12.351  & 11.803  & 12.914   & 12.147   & 11.526  & 12.827   \\
      $\rho_\text{max}$              & 0.806   & 0.719   & 0.898    & 0.871    & 0.767   & 0.968    \\
      $\epsilon_\text{max}$          & 958.332 & 832.054 & 1094.927 & 1053.422 & 898.290 & 1211.631 \\
      $c_{s,\text{max}}^2$           & 0.513   & 0.320   & 0.700    & 0.564    & 0.340   & 0.795    \\
      $M_\text{Q,max}$               & 0.126   & 0.032   & 0.447    & 1.176    & 0.777   & 1.555    \\
      $R_\text{Q,max}$               & 3.409   & 2.116   & 5.298    & 8.069    & 7.026   & 8.952    \\
      $R_{1.4 \text{M}_\odot}$       & 12.964  & 12.725  & 13.298   & 13.448   & 13.193  & 13.698   \\
      $M_\text{trans}$               & 2.147   & 2.012   & 2.266    & 1.634    & 1.253   & 1.940    \\
      \hline
    \end{tabular}
    \caption{The median and the 90\% CI of the following quantities:
    {the phase transition
    density ($\rho_\text{trans}$, in fm$^{-3}$),
    pressure ($P_\text{trans}$, in MeV/fm$^3$),
    hadronic energy density ($\epsilon_\text{H,trans}$, in MeV/fm$^3$) and
    quark energy ensity ($\epsilon_\text{Q,trans}$, in MeV/fm$^3$); 
    measure of the strength of the phase transition
    ($\Delta \epsilon_\text{trans} = \epsilon_\text{Q,trans} -\epsilon_\text{H,trans}$, in MeV/fm$^3$);
    onset of the quark strange ($\rho_s$, in $\fm$);
    fraction of the quark strange in the maximum mass NS ($Y_{s,\text{max}}$);
    maximum mass ($M_\text{max}$, in M$_\odot$) and
    radius ($R_\text{max}$, in km),
    density ($\rho_\text{max}$, in fm$^{-3}$),
    energy density ($\epsilon_\text{max}$, in MeV/fm$^3$),
    speed of sound ($c_{s,\text{max}}^2$)
    quark core mass ($M_\text{Q,max}$, in M$_\odot$) and
    quark core radius ($R_\text{Q,max}$, in km) at the maximum NS mass;
    radius of the $1.4 \text{M}_\odot$ ($R_{1.4 \text{M}_\odot}$, in km);
    mass of stars with central pressure equal to $P_\text{trans}$ ($M_\text{trans}$, in M$_\odot$).
    }}
    \label{tab:NS_prop}
  \end{table*}

    \subsection{Discussion of the properties of NS}
    \label{sec:results_ns}

\begin{figure}[ht]
        \centering
        \includegraphics[width=.9\linewidth]{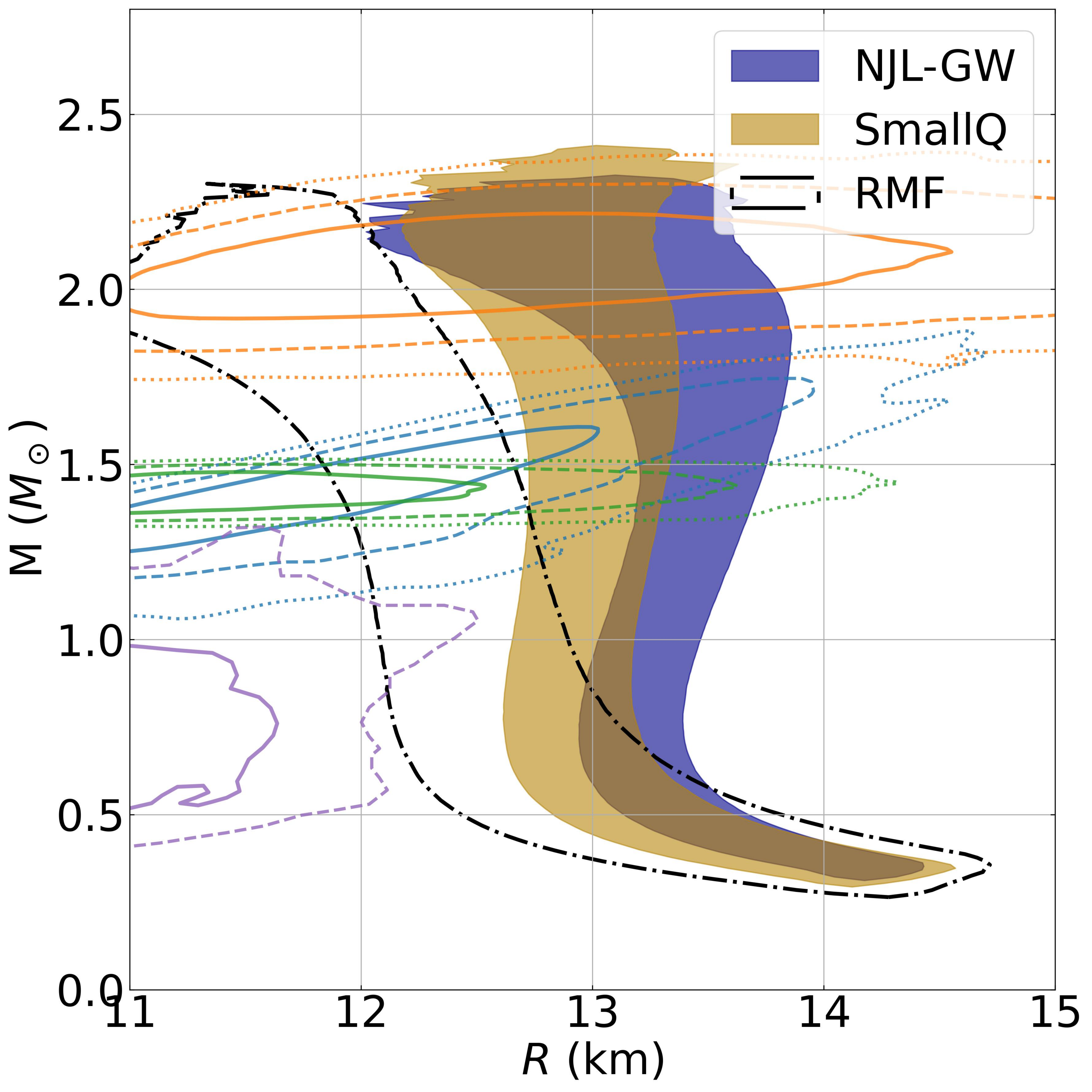}
        \caption{Mass-radius diagram of the 90\% CI for the NJL-GW (purple), SmallQ (yellow) and RMF (black dash-dot) sets.
            The NICER observational data
            PSR J0030+0451 \cite{miller_2019,vinciguerra_2024} (blue),
            PSR J0740+6620 \cite{miller_2021,salmi_2024} (orange),
            PSR J0437+4715 \cite{choudhury_2024} (green), and
            and the HESS data J1731-347 \cite{doroshenko_2022} (purple)
            with $1 \sigma$, $2 \sigma$, $3 \sigma$
            are shown with solid, dashed and dotted lines, respectively.
            }
        \label{fig:tov}
    \end{figure}

\begin{figure}[ht]
        \centering
        \includegraphics[width=.9\linewidth]{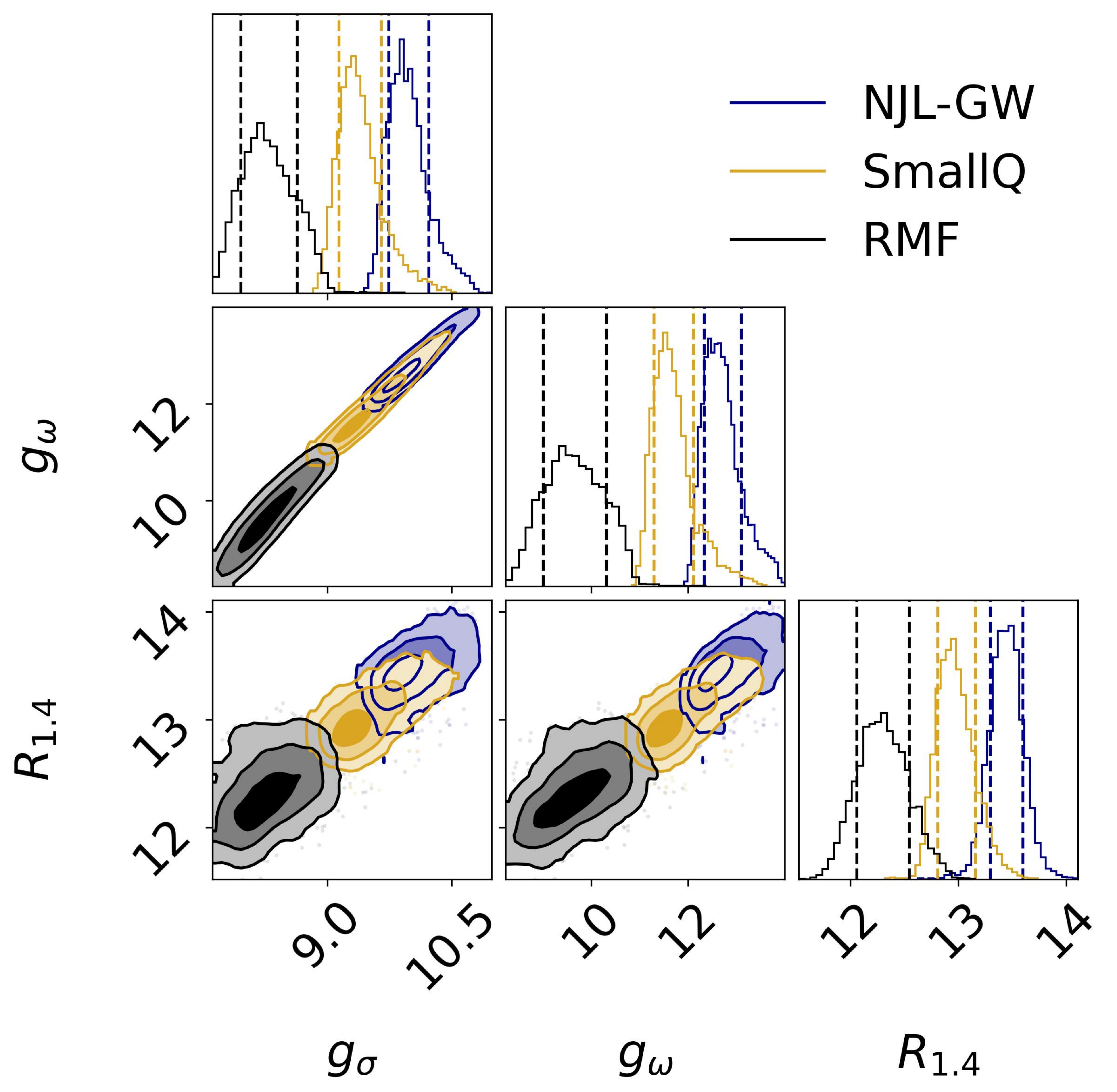}
        \caption{Corner plot of $g_\sigma$, $g_\omega$ and $R_{1.4}$ of the sets
            NJL-GW (purple), SmallQ (yellow) and RMF (black).}
        \label{fig:cornerR14}
    \end{figure}

The EOS, pressure as a function of the energy density, is the function needed to integrate the Tolman-Oppenheimer-Volkov (TOV)  equations and determine the NS properties.
    % Figure M x R + corner R14
	Figure \ref{fig:tov} shows the 90\% CIs of the mass-radius diagram
	for the NJL-GW and the SmallQ sets.
	In this figure, we also show the results of the RMF set,
	which contains only nucleonic matter.
	The NICER observational data of the
	PSR J0030+0451 \cite{miller_2019,vinciguerra_2024} (blue),
	PSR J0740+6620 \cite{miller_2021,salmi_2024} (orange),
	PSR J0437+4715 \cite{choudhury_2024} (green), and
	the HESS J1731-347 \cite{doroshenko_2022} (purple)
	with $1 \sigma$, $2 \sigma$, and $3 \sigma$
	are represented by solid, dashed, and dotted lines, respectively.
	The SmallQ set predicts smaller radii
	and can better describe the observational data than the NJL-GW set.
    This difference occurs because the SmallQ EoS is softer than the NJL-GW  EoS in the hadron phase.
    A larger $g_\omega$ parameter stiffens the EoS. This coupling 
    is correlated with the radius of the $1.4 M_\odot$ NSs ($R_{1.4}$),
    as shown in Figure \ref{fig:cornerR14}. The corner plot  also shows a strong correlation with the $g_\sigma$ coupling which is
    imposed by the binding energy at saturation.

\begin{figure}[ht]
        \centering
        \includegraphics[width=.9\linewidth]{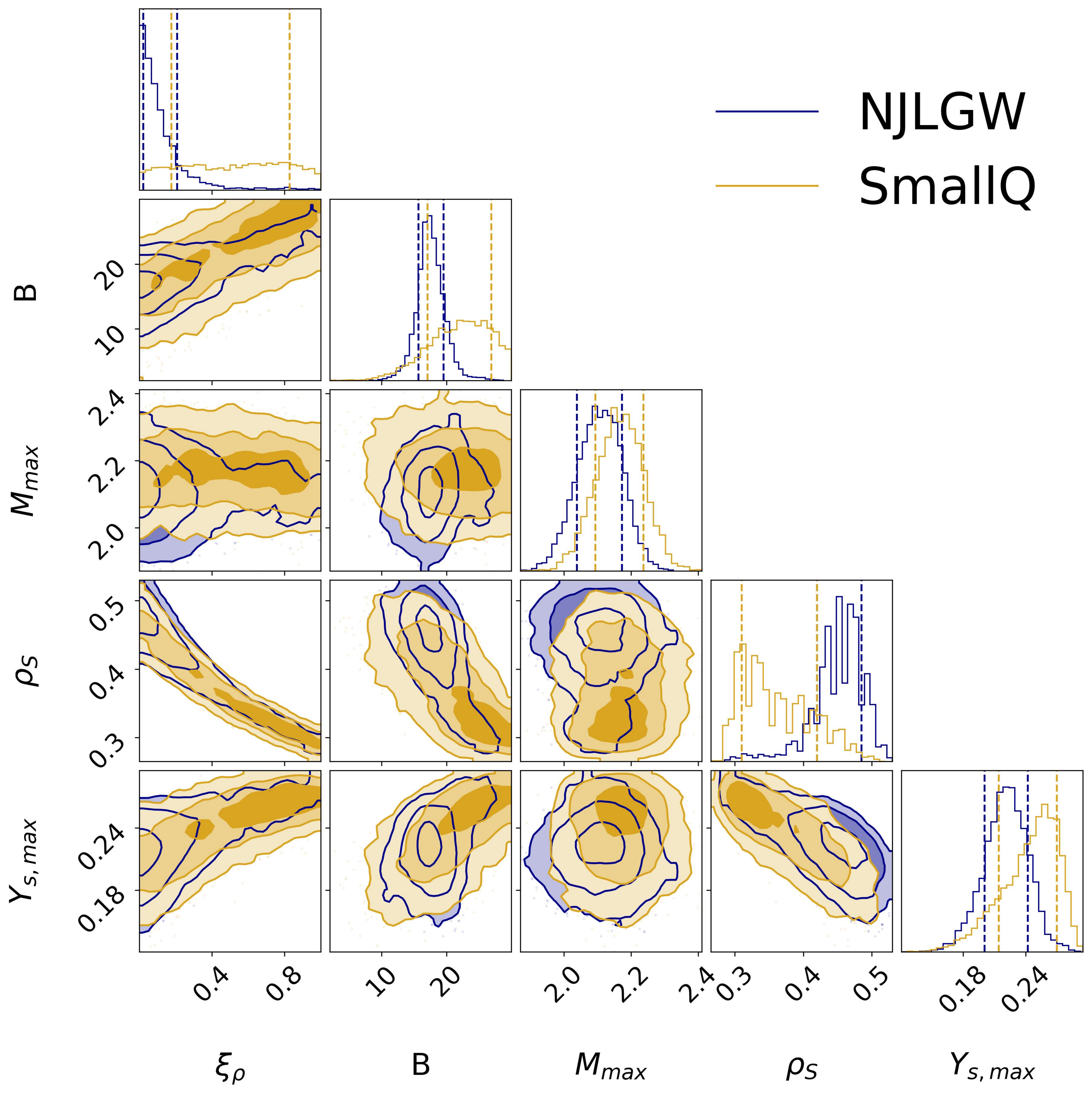}
        \caption{Corner plot of $\xi_\varrho$, $B$, $\Mmax$, $\rho_s$ and $Y_{s,\text{max}}$
            for the sets NJL-GW (purple) and SmallQ (yellow).}
        \label{fig:corner_strange}
    \end{figure}

\begin{figure}[ht]
        \centering
        \includegraphics[width=.9\linewidth]{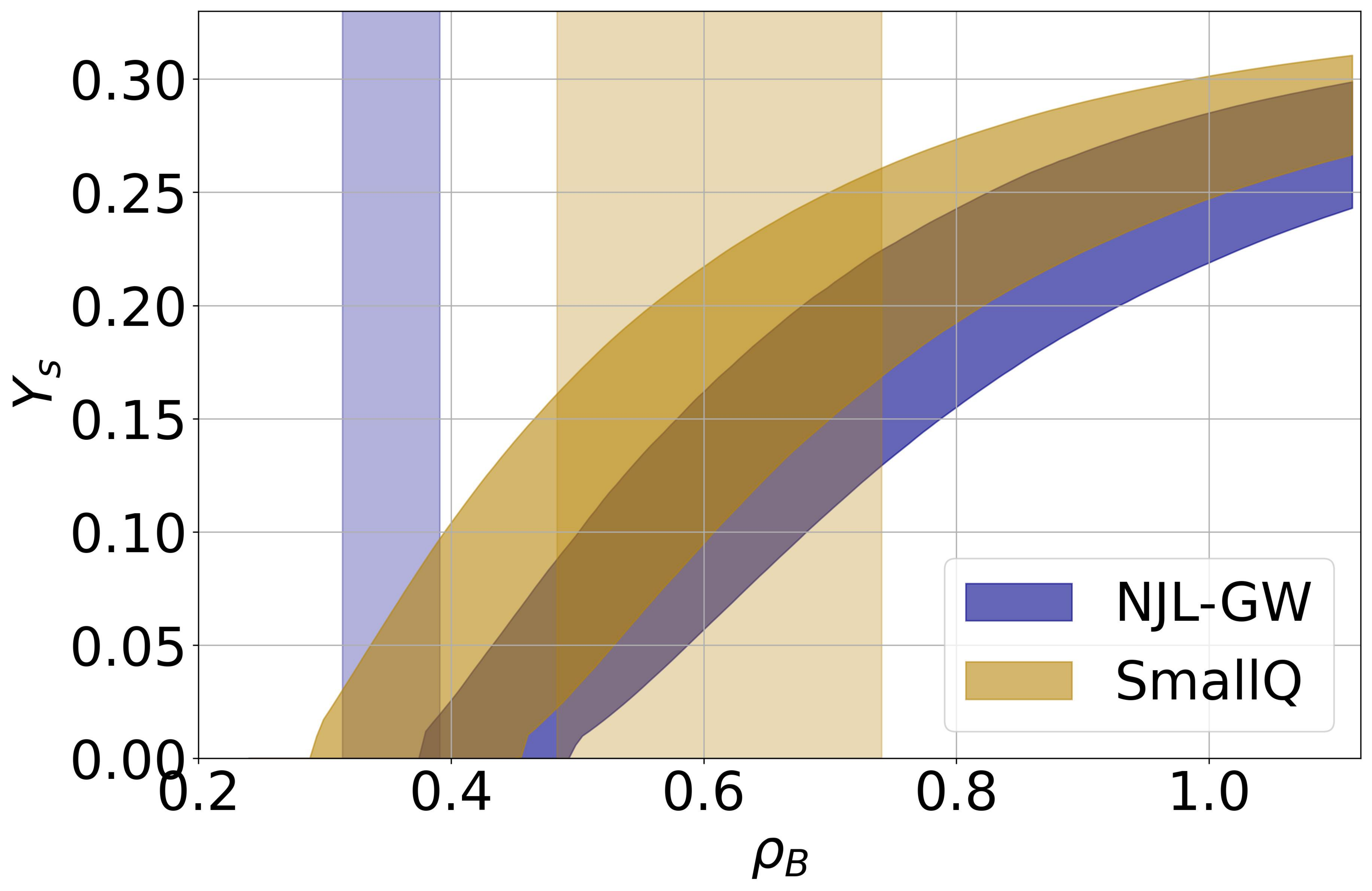}
        \caption{Strange quark fraction, 90\% CI, as a function of the baryonic density for SmallQ and NJL-GW sets.
        Vertical bands identify the 90\% CI for the phase transition density.}
        \label{fig:frac_strange}
    \end{figure}

    % corner strange
    In Figure \ref{fig:corner_strange}, we show the corner plot of
    the quark parameters $\xi_\rho$ and $B$,
    and the properties of maximum mass NS ($\Mmax$),
    the density at which the quark strange appears ($\rho_s$), and
    the quark strange fraction in the maximum mass NS ($Y_{s,\text{max}})$.
    The other parameters of the quark phase do not change significantly.
    We notice that the distribution of the $\xi_\rho$ parameter
    has a peak at $0$ in the NJL-GW set.
    However, in the SmallQ set, this parameter is uniformly distributed.
    The $\xi_\rho$ has a strong correlation with the $\rho_s$:
    increasing the $\xi_\rho$ value,
    the onset of the quark strange occurs at lower densities.
    As a consequence, the EoS becomes softer, and the phase transition density increases.
    This effect of $\xi_\rho$ can also be seen in \cite{ferreira_2021}.
    The EOS of quark matter must be stiff  to contain a large quark core.  Therefore, it disfavors the onset of the s-quark, as in the NJL-GW set. Having relaxed this constraint, there is more freedom concerning the quark matter composition, and it is interesting to verify that more symmetric matter is favored with a larger s-quark fraction. In Fig. \ref{fig:frac_strange}, the 90\% CI of the s-quark fraction is plotted as a function of the baryonic density. Except for the larger density, the s-quark fraction is generally $\gtrsim$ 50\% larger for the SmallQ model.

    The softening of the SmallQ EoS is compensated by an increase in the $B$ values, as this term stiffens the EoS and increases the $\Mmax$ values. Indeed, the corner plot shows a clear  correlation between the $\xi_\rho$ and $B$.  This explains that, although $\rho_s$ decreases within the SmallQ set, increasing the values of the fraction of strangeness $Y_{s,\text{max}}$,  the $\Mmax$ still increases in relation to the NJL-GW set.

\begin{figure}[ht]
        \centering
        \includegraphics[width=.9\linewidth]{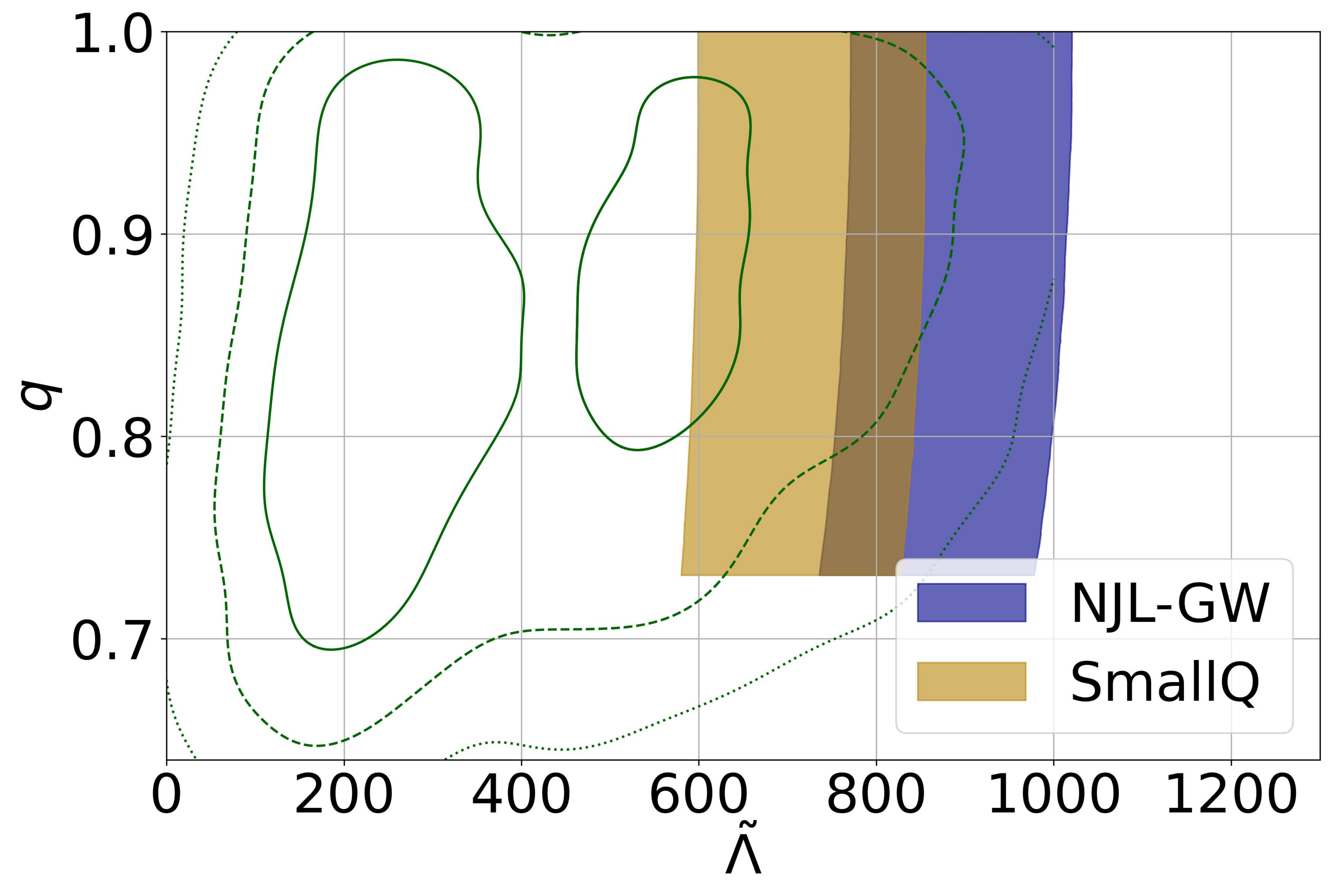}
        \caption{Binary mass ratio as a function of tidal deformability
            for the 90\% CIs of the NJL-GW (purple) and the SmallQ (yellow) sets.
            The LIGO-Virgo collaboration data GW170817
            \cite{ligovirgo_2018}
            is represented by the green contours,
            with $50 \%$ CI (solid), $90 \%$ CI (dashed) and $99 \%$ CI (dotted).}
        \label{fig:LT}
    \end{figure}

    % Figure tidal
    Figure \ref{fig:LT} shows the binary mass ratio $q = M_2/M_1 < 1$
    as a function of the effective tidal deformability,
    defined as
    \begin{equation}
        \tilde{\Lambda}
        = 
        \frac{16}{13} \frac{(12 q +1) \Lambda_1 +(12 +q) q^4 \Lambda_2}{(1 +q)^5},
    \end{equation}
    where $\Lambda_i$ and $M_i$ are the tidal deformability and the mass
    of the i-th NS of the binary system, respectively.
    The $\tilde{\Lambda}$ is calculated for $0.73 < q < 1$ for
    a fixed chirp mass of
    $\Mchirp = (M_1 M_2)^{3/5}/(M_1 +M_2)^{1/5} = 1.186 \Msun$.
    This value corresponds to the LIGO-Virgo collaboration data
    from GW170817 \cite{ligovirgo_2018},
    which is represented in Figure \ref{fig:LT} for
    50\%, 90\% and 99\% CIs in green solid, dashed and dotted contours, respectively.
    In this figure, the NJL-GW set struggles to satisfy the GW170817 data,
    lying mostly within the 99\% CI.
    However, the SmallQ model provides a better description of this data and
    achieves the 50\% CI contour.

    % Table slope
    In \cite{ferreira_2024}, analyze the effect of  the slope $dM/dR$ on NS properties, considering  a model-independent EoS. Properties such as
   maximum mass, radius, and the position of the $c_s^2$ peak, where considered. In \cite{ferreira_2025}, a similar analysis was carried out with EOS sets based on microscopic descriptions. The RMF model was used to compare the slope  $dM/dR$ of nucleonic and hyperonic EoS.  EOS with hyperons tend to have a positive slope at low masses.
    However, %a notable difference is that 
    all the hyperonic EoS have a negative slope, at $M = 1.8 \Msun$.

    In Table \ref{tab:slope_MR},
    we show the sign of the slope $dM/dR$ at $M = 1.2, \; 1.4, \; 1.6 \; 1.8 \Msun$
    for the NJL-GW and SmallQ hybrid sets and the RMF hadronic set from \cite{albino_2026}.
    The hadronic set has a predominantly negative slope at all masses.
    For the hybrid sets, the positive slope is predominant at $M = 1.2, \; 1.4 \Msun$.
    At $1.6 \Msun$, $47 \%$ ($52 \%$) of the EoS in the SmallQ set have a positive (negative) slope,
    while $62 \%$ of the EoS in the  NJL-GW set has a positive slope.
    At $1.8 \Msun$, the sign is inverted, and $78 \%$ ($89 \%$) of the EoS have a negative slope. Notice, however, that there is still a non negligible fraction of hybrid EOS with a positive slope, while for the hyperonic set no EOS is showing a positive slope at  $1.8 \Msun$.
    Thus, a positive slope at low masses could indicate the presence of quark matter inside NS. Moreover, a positive slope at $1.8 \Msun$ could certainly be a signature of the presence of  quark matter. This behavior was also observed in our previous work \cite{albino_2026}.
    %
    %In \cite{ferreira_2025}, the RMF model was used to compare the slope of hadronic and hyperonic EoS. As with hybrid EoS, equations with hyperons tend to have a positive slope at low masses. However, a notable difference is that all the hyperonic EoS have a negative slope, at $M = 1.8 \Msun$.
    %
	\begin{table*}[ht]
		\centering
		\setlength{\tabcolsep}{12.pt}
		\renewcommand{\arraystretch}{1.2}
		\begin{tabular}{cccccccccc}
		  \hline \hline
		  $M$ & \multicolumn{2}{c}{$1.2 \text{M}_\odot$} & \multicolumn{2}{c}{$1.4 \text{M}_\odot$}
		    & \multicolumn{2}{c}{$1.6 \text{M}_\odot$} & \multicolumn{2}{c}{$1.8 \text{M}_\odot$} 
            & Ref. \\
		  $dM/dR$ & + & - & + & - & + & - & + & - & \\
		  \hline
		  NJL-GW    &  7268 &   253 &  6656 &   865 & 4699 & 2822 & 1653 &  5868 & \cite{albino_2026}   \\
		  SmallQ    &  5501 &   465 &  4913 &  1053 & 2817 & 3149 &  643 &  5323 & this work            \\
		  RMF       &   822 &  5215 &   175 &  5862 &   70 & 5967 &   38 &  5999 & \cite{albino_2026}   \\
        nucleonic &  4511 & 13025 &  1514 & 16023 &   -- &   -- &   47 & 17537 & \cite{ferreira_2025} \\
        hyperonic & 16501 &    95 & 15953 &   193 &   -- &   -- &    0 & 16146 & \cite{ferreira_2025} \\
		  \hline
		\end{tabular}
		\caption{Number of EOSs with positive/negative slope in the mass-radius diagram at
		    $M = 1.2, 1.4, 1.6, 1.8 \text{M}_\odot$ for the SmallQ set,
            NJL-GW and RMF sets from \cite{albino_2026}
            and nucleonic and hyperonic sets from \cite{ferreira_2025}.}
		\label{tab:slope_MR}
	\end{table*}

    \subsection{Discussion on the Phase Transition}

\begin{figure}[ht]
        \centering
        \includegraphics[width=.9\linewidth]{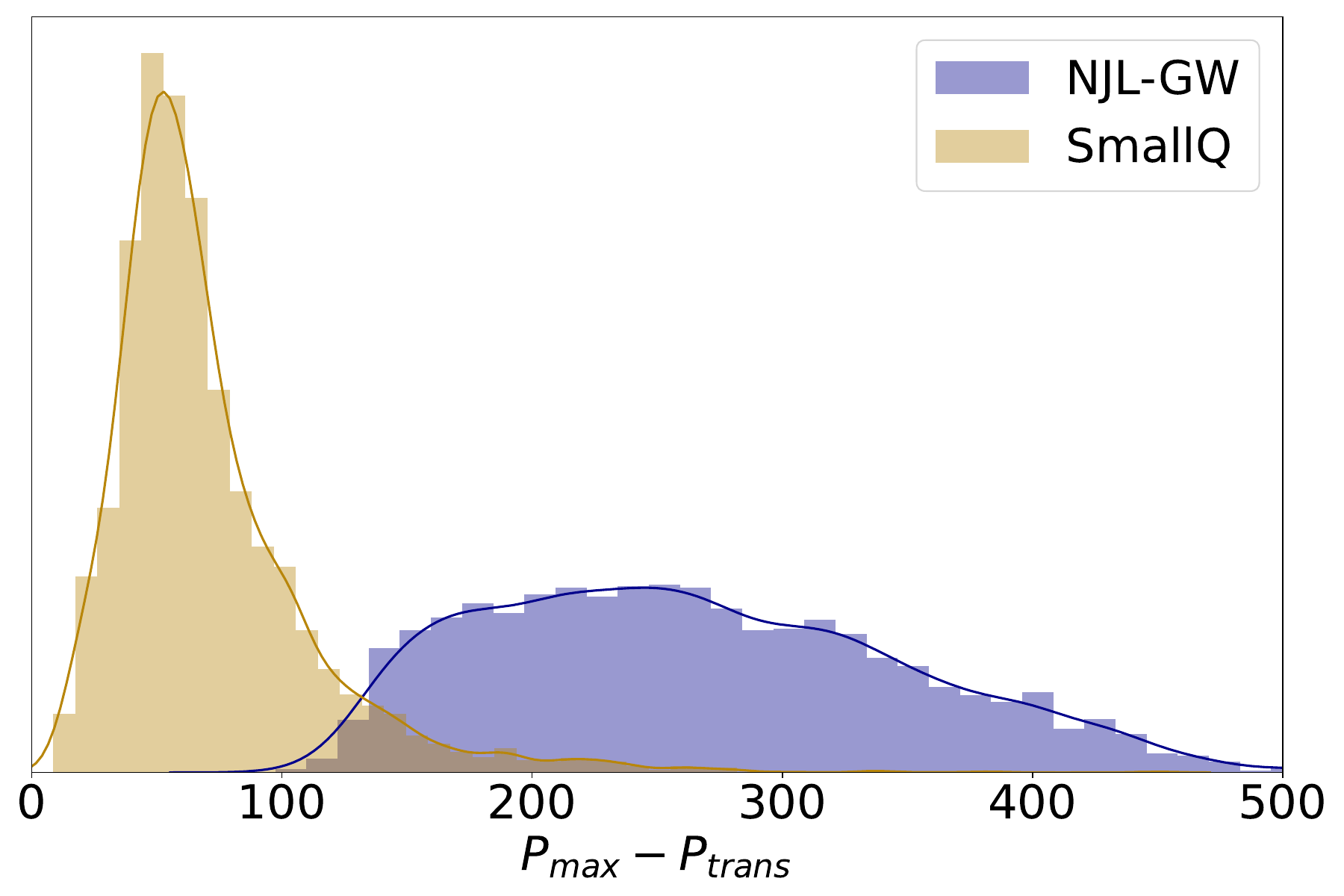}
        \caption{Distribution of $P_{max} -P_{trans}$ value for the NJL-GW (purple) and SmallQ (yellow) sets.}
        \label{fig:PmaxPtrans}
    \end{figure}

    % Distribution $P_max -P_trans$
    One of the main differences between the NJL-GW and SmallQ sets is the
    restriction of the $P_\text{max} -P_\text{trans}$ value.
    The $P_\text{max}$ quantity is the central pressure of the maximum mass NS,
    while the $P_\text{trans}$ is the pressure of the phase transition.
    The difference $P_\text{max} -P_\text{trans}$ is constrained in the Bayesian inference
    as a means of controlling the size of the quark core.
    For the NJL-GW set, we impose that $P_\text{max} -P_\text{trans} > 100 \MeVfm$,
    resulting in NSs with large quark cores.
    For the SmallQ set, however, we relaxed this restriction and used $P_\text{max} -P_\text{trans} > 5 \MeVfm$.
    This means that this set still only includes hybrid stars, but allows for any size of quark core.
    In Figure \ref{fig:PmaxPtrans}, the distribution of $P_\text{max} -P_\text{trans}$ are shown
    for the NJL-GW (purple) and SmallQ (yellow) sets.
    We notice that the SmallQ distribution is narrower than the NJL-GW one
    and the curve is concentrated at lower values.

\begin{figure}[ht]
        \centering
        \includegraphics[width=.9\linewidth]{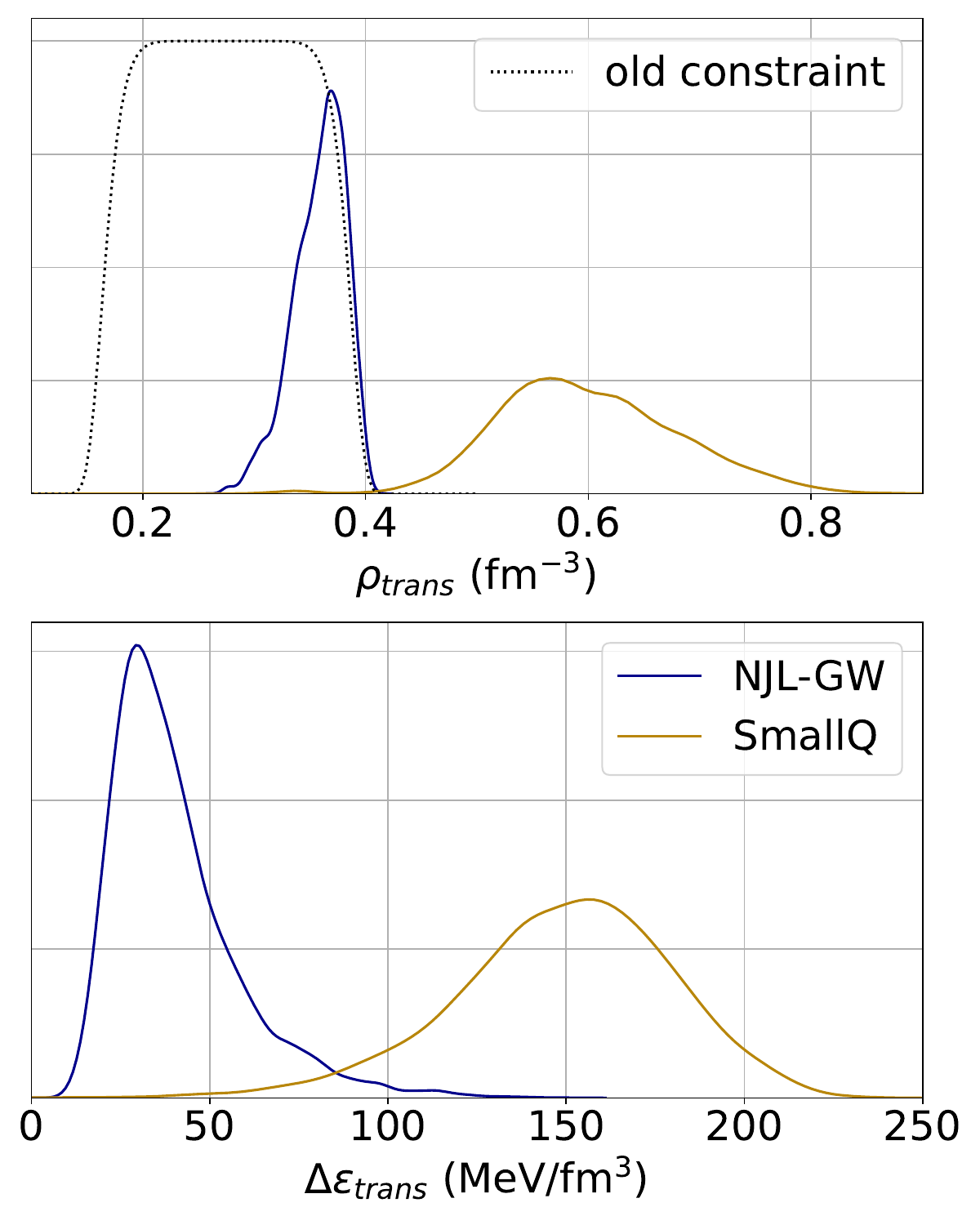}
        \caption{Distribution of phase transition density (top) and
            jump in the energy transition (bottom) for the NJL-GW (purple) and SmallQ (yellow) sets.
            The constraint in $\rho_\text{trans}$ imposed in the NJL-GW set is represented
            by the black dotted line.
            }
        \label{fig:distphT}
    \end{figure}

    % Distribution phase transition
    Another difference between the NJL-GW and SmallQ sets is the constraint for the
    phase transition density ($\rho_\text{trans}$).
    For the NJL-GW set, we imposed that the $\rho_\text{trans}$ must have values between
    $\sim 0.16$ and $0.40 \fm$.
    On the other hand, for the SmallQ set,
    we did not constraint the $\rho_\text{trans}$.
    The top panel of Figure \ref{fig:distphT} shows the distribution of the phase transition density
    in purple and yellow curves for the NJL-GW and the SmallQ sets, respectively.
    The constraint imposed for the NJL-GW set is shown in black dotted line.
    Notably, while the values of $\rho_\text{trans}$ for the NJL-GW set
    remains small, satisfying the constraint,
    the SmallQ results in larger values of phase transition density,
    with $\rho_\text{trans}$ between $0.484$ and $0.741 \fm$ (90\% CI, see Table \ref{tab:NS_prop}).
    This also affects the jump in the energy density at the phase transition ($\Delta \epsilon_\text{trans}$).
    The distribution of this quantity is shown in the bottom panel of Figure \ref{fig:distphT},
    in which we see that the SmallQ set has larger values for $\Delta \epsilon_\text{trans}$
    than the NJL-GW set, indicating that the former set has a stronger phase transition.
    
    %[Fig. 9 and 10 show the whole problem. The hadronic RMF scenario wins in likelihood given the observations. Assuming and enforcing the quarks, without any other knowledge/constraint, the observations likelihood force a tiny QC so we could have the largest M-R portion possible in the most likelihood scenario which is the hadronic star --- transition density increases. Do you agree?]
    Given the constraints, in particular, the astrophysical observations, the hadronic EOS are favored with respect to the EOS with a large quark core. Giving the freedom to the model to determine the optimal size of the quark core, the model chooses a small quark core which defines a behavior similar to the one of the hadronic EOS. The transition to a quark phase within an NJL description is dictated by the density at which chiral restoration occurs, which occurs above twice times the saturation density.  A model that predicts a restoration of chiral symmetry at smaller densities would probably favor larger quark cores. This will be investigated in the future.

\begin{figure}[ht]
        \centering
        \includegraphics[width=.9\linewidth]{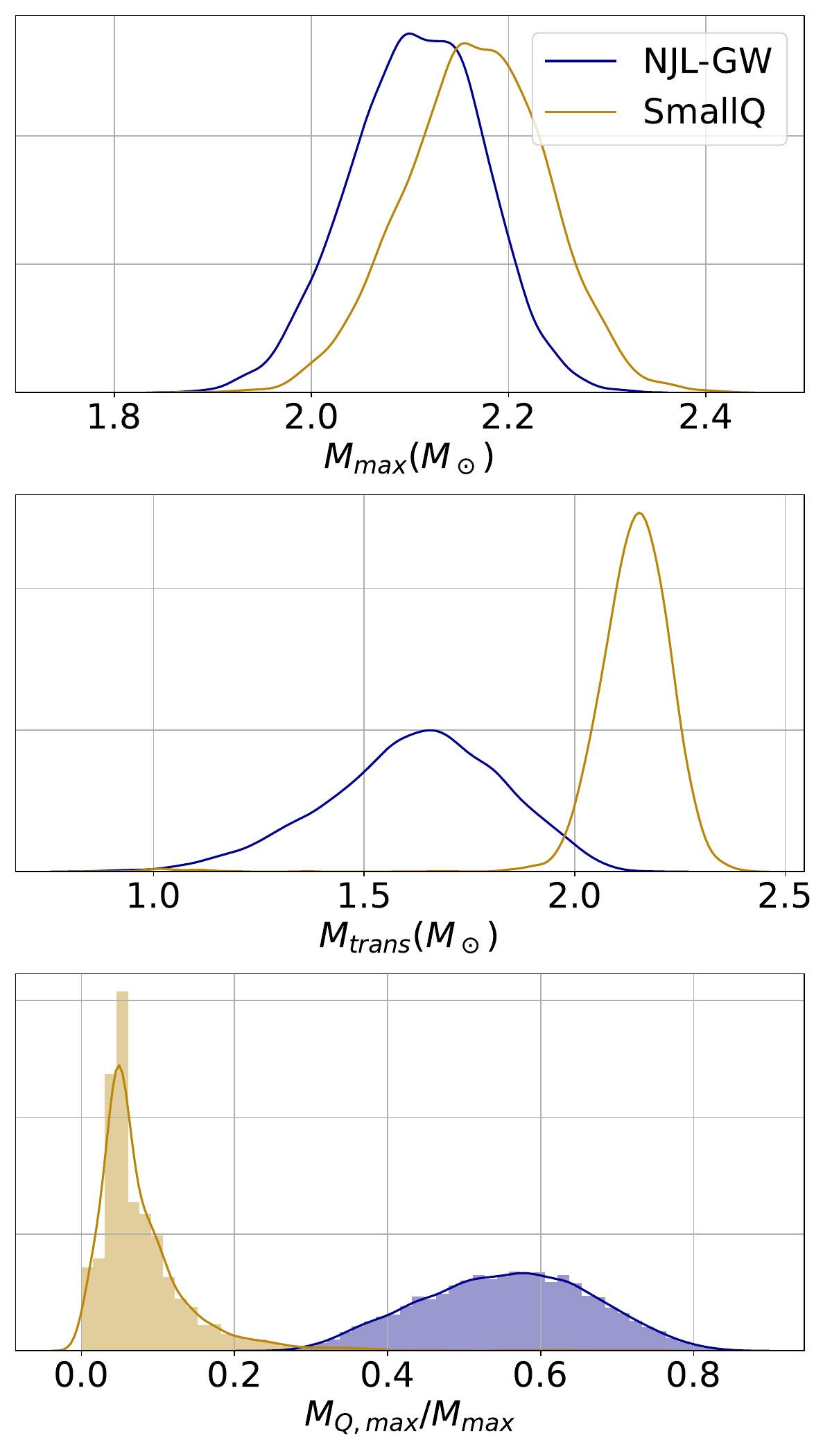}
        \caption{Distribution of the maximum mass ($\Mmax$),
            phase transition mass ($\Mtrans$) and
            the fraction of the quark core mass ($M_{Q,\text{max}}$) by
            the total mass in the maximum mass NS.}
        \label{fig:distM}
    \end{figure}

    % Distribution NS mass
    Figure \ref{fig:distM} shows the distributions of
    the maximum mass NS ($\Mmax$),
    the mass in which the central density is equal to the phase transition density ($\Mtrans$),
    and the fraction of the quark core mass by the total mass of the maximum mass NS ($M_\text{Q,max}/\Mmax$).
    The SmallQ set has slightly larger values for the maximum mass NS than the NJL-GW set.
    However, both sets result in different $\Mtrans$ distributions. 
    The NJL-GW set results in smaller values of $\Mtrans$ with a wider distribution.
    This set indicates that NS with $M = 1.634 \Msun$ (median value -- see Table \ref{tab:NS_prop})
    can have quark matter in their core.
    On the other hand, the SmallQ set has a narrower distribution with larger values for the $\Mtrans$.
    In Table \ref{tab:NS_prop}, the minimum of the 90\% CI is $\Mtrans = 2.012 \Msun$ (Table \ref{tab:NS_prop}),
    indicating that NS with a quark core must have at least this mass.
    The sets also show a different behavior for the distribution of the quark core fraction in the maximum mass NS.
    The fraction of the quark mass for the NJL-GW set can vary from $\sim 30 \%$ to $80 \%$,
    while for the SmallQ set, most part of the distribution lies in $< 10 \%$.
    Notably, by relaxing the constraints, the Bayesian inference indicates a preference
    for NSs with smaller quark core.

    \subsection{Nuclear Matter Properties}

\begin{figure*}[ht]
        \centering
        \includegraphics[width=.9\linewidth]{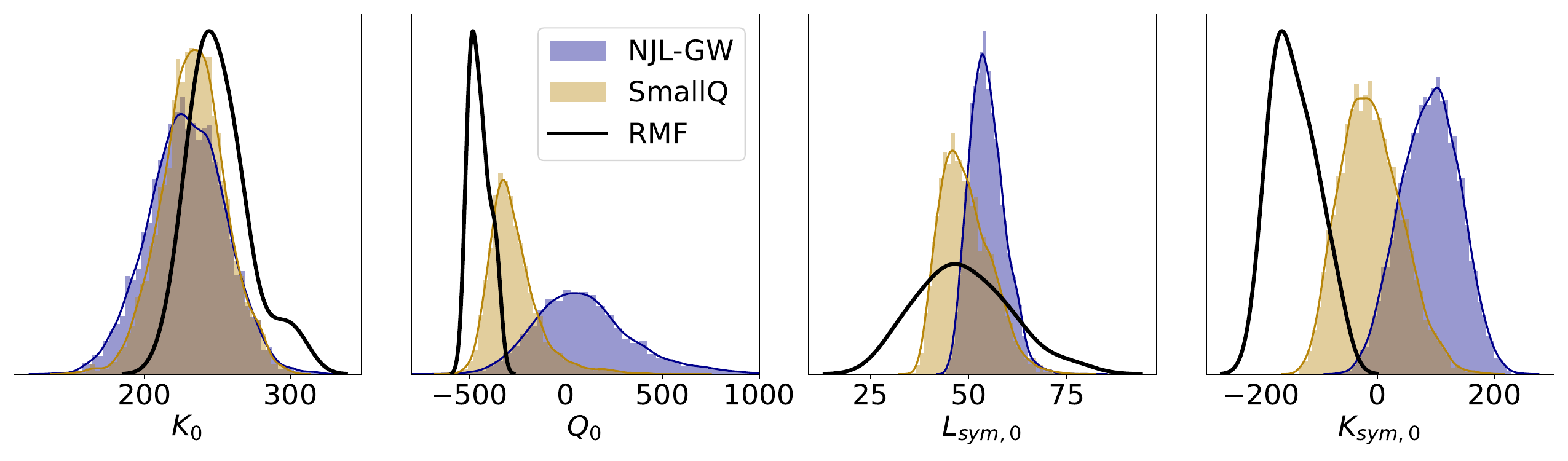}
        \caption{Nuclear matter properties distributions for the
            NJL-GW (purple), SmallQ (yellow) and RMF (black) sets.}
        \label{fig:nmp}
    \end{figure*}

    In this section, we discuss the nuclear matter properties (NMP).
    Table \ref{tab:NMP_prop} presents the 90 \% CIs of all the NMP quantities for the SmallQ, NJL-GW and RMF sets. These quantities are defined in Eq.~(\ref{eq:nmp_expansion}).
    Figure \ref{fig:nmp} shows the distributions of $K_0$, $Q_0$, $L_{\text{sym},0}$ and $K_{\text{sym},0}$
    for the NJL-GW (purple), SmallQ (yellow) and RMF (black) sets.
    Since the distributions of $\rho_0$, EA and $J_{\text{sym},0}$ are constrained in the Bayesian inference,
    they do not show a significant difference in the plots and are therefore omitted here.
    Analyzing the figure, the SmallQ set distributions tend to lie between the NJL-GW and RMF results.
    This effect is because the phase transition of the SmallQ set
    occurs at larger densities than the NJL-GW set, and therefore, the effect of the presence of quarks at high densities is smaller.
    Since the SmallQ EoS contains more nucleonic matter,
    the distributions of this set shift towards the RMF ones.
    Although $K_0$ is constrained, there are some differences in these distributions.
    While the NJL-GW and SmallQ sets satisfy the constraint,
    the RMF shifts to larger values. This can be understood in the following way: the nucleonic RMF EOS has to be stiff to reach 2 $\Msun$, and the skewness $Q_0$ should not be too large in order to satisfy pQCD constraints. In the hybrid EOS, the hadron phase is more compatible with  $K_0$ constraints, however the skewness is much larger to compensate. In addition,
    the stiffness of the quark phase can help reaching 2$\Msun$. In this case, the transition to quark matter helps satisfying pQCD constraints even with a large $Q_0$.
    We observe that SmallQ has slightly larger values of $K_0$ than NJL-GW
    for the 90 \% CI minimum.  However, the 90 \% CI maximum remains the same.
    For the $L_{\text{sym},0}$,
    although the RMF distribution is wider than that of the hybrid sets,
    we notice that the median value is similar for all of them.
    These values are also consistent with
    $L_{\text{sym},0} = 58.7 \pm 28.1 \MeV$,
    obtained by averaging different results from the literature in \cite{oertel_2017}.

    In the $K_{\text{sym},0}$ plot,
    the RMF distribution implies that this quantity has a negative sign,
    in agreement with \cite{li_2021},
    in which the authors obtained $K_{\text{sym},0} = -107 \pm 88 \MeV$
    by averaging the result of 16 studies.
    However, positive values of $K_{\text{sym},0}$, as obtained from NJL-GW,
    are compatible with some results from \cite{reed_2024},
    in which the CREX and PREX-II data were considered.
    The SmallQ set accepts both positive and negative values of $K_{\text{sym},0}$.

  \begin{table*}[ht]
    \centering
    \setlength{\tabcolsep}{8.pt}
    \renewcommand{\arraystretch}{1.2}
    \begin{tabular}{cccccccccc}
      \hline \hline
      & \multicolumn{3}{c}{Set SmallQ} & \multicolumn{3}{c}{Set NJL-GW} & \multicolumn{3}{c}{Set RMF} \\
      quant & median & min & max & median & min & max & median & min & max \\
      \hline
      $\rho_0$         & 0.159    & 0.158    & 0.161    & 0.159    & 0.158    & 0.161     & 0.159    & 0.158    & 0.161   \\
      EA               & -16.001  & -16.033  & -15.967  & -16.000  & -16.032  & -15.968   & -16.000  & -16.032  & -15.968 \\
      $K_0$            & 233.810  & 197.247  & 270.421  & 228.829  & 185.733  & 271.410   & 250.940  & 228.158  & 283.916 \\
      $Q_0$            & -296.730 & -434.109 & -23.609  & 71.412   & -273.811 & 573.546   & -445.834 & -510.475 & -346.521\\
      $J_\text{sym,0}$ & 31.640   & 29.217   & 34.296   & 31.149   & 28.706   & 33.708    & 32.138   & 29.678   & 34.717  \\
      $L_\text{sym,0}$ & 48.425   & 40.484   & 60.597   & 54.256   & 48.117   & 62.705    & 46.898   & 34.300   & 66.925  \\
      $K_\text{sym,0}$ & -14.559  & -88.959  & 79.321   & 91.128   & 6.448    & 169.655   & -140.665 & -184.537 & -69.975 \\
      $Q_\text{sym,0}$ & 1264.450 & 784.071  & 1551.287 & 1002.040 & 202.071  & 1506.580  & 1136.710 & 407.590  & 1563.186\\
      \hline
    \end{tabular}
    \caption{Median and 90\% CI of the NMP (defined in Eq.~(\ref{eq:nmp_expansion})) for the SmallQ, NJL-GW and RMF sets:
saturation density $\rho_0$ (fm$^{-3}$), binding energy per nucleon $EA$, incompressibility $K_0$, skewness $Q_0$,
symmetry energy $J_\text{sym,0}$, and its slope $L_\text{sym,0}$, curvature $K_\text{sym,0}$ and skewness $Q_\text{sym,0}$ (all in MeV).}
    \label{tab:NMP_prop}
  \end{table*}

\section{Conclusion}
\label{sec:conclusion}

  In this work, we compare two sets of hybrid EoS,
  one that reproduces NSs with large quark cores,
  and the other, with small quark cores.
  The EoS are obtained using the Relativistic Mean Field (RMF) model
  with non-linear meson interaction terms for the hadron phase,
  and the Nambu--Jona-Lasinio (NJL) model
  with four- and eight-quark interaction terms for the quark phase.
  We apply the Maxwell construction to reproduce the phase transition.
  We use Bayesian inference to impose
  nuclear matter properties,
  pure neutron matter results from chEFT,
  the pQCD constraint at 7 $\rho_0$,
  as developed in \cite{gorda_2023,komoltsev_2023},
  the X-ray NICER data of
  J0030-0451 \cite{vinciguerra_2024,miller_2019},
  J0740-6630 \cite{salmi_2024,miller_2021},
  J0437-4715 \cite{choudhury_2024,reardon_2024}, and
  the gravitational wave data from
  GW170817 \cite{abbott_2018}.
  We also apply some constraints to ensure a phase transition.
  For the NJL-GW set,
  we impose that the quark core of the $\NSmax$ must be large,
  while for the SmallQ set,
  we only impose that the $\NSmax$ must have a quark core,
  regardless of its size.

  Due to the early quark-hadron phase transition in the NJL-GW set,
  the hadron phase is stiffer compared with the SmallQ set.
  As a consequence, the SmallQ set reproduces NSs with smaller radius and tidal deformability,
  which are more compatible with the NICER data and the GW170817 data.
  This also affects the composition of the quark phase,
  resulting in a larger quark strange fraction for the SmallQ set.
  Also, the phase transition density is larger than the onset of the quark strange density,
  in contrast with the NJL-GW set.
  This implies larger values of the speed of sound squared at
  intermediate densities for the SmallQ set.
  Furthermore,
  the SmallQ set is more compatible with the 90\% CI agnostic model of \cite{altiparmak_2022},
  and also with the conditions of the quark phase behavior
  ($\gamma < 1.75$ and $d_c < 0.2$)
  set for the agnostic models in \cite{annala_2020,annala_2023}.

  NJL-GW set indicates the presence of quark matter inside the $2 \Msun$ NS,
  while the SmallQ set indicates that these stars may contain little to no quark matter.
  Both sets have a predominantly positive mass-radius derivative for low masses.
  This result was already analyzed in the previous work \cite{albino_2026}
  and in \cite{ferreira_2025},
  which also indicate $dM/dR > 0$ for exotic (quark and hyperon) matter
  and $dM/dR < 0$ for hadron matter. In addition, there is still a non negligible fraction of EOS that present a positive slope for 1.8 $\Msun$ stars, while for nucleonic stars this fraction is negligible and for hyperonic stars it was shown that for masses of this order or larger no EOS would have a positive slope.
  Observational data with more accurate radius values may favor the presence of exotic matter
  if it confirms that the $2 \Msun$ NSs, such as the J0740+6620,
  has larger radius than the $1.4 \Msun$ NSs (J0030+0451 and J0437-4715).

\section*{Data Availability}
The final posterior of the model parameters,
the equation of states and the TOV results of the
SmallQ set can be obtained from the link
\cite{albino_2026_23471790}.
The results for NJL-GW and RMF sets are available at
\cite{albino_2025_17534438}.

\appendix

\section{Corner Plots of the Posterior}

\begin{figure*}[ht]
      \centering
      \begin{tabular}{cc}
      \includegraphics[width=.4\linewidth]{Images/fig13a.pdf}
      &
      \includegraphics[width=.4\linewidth]{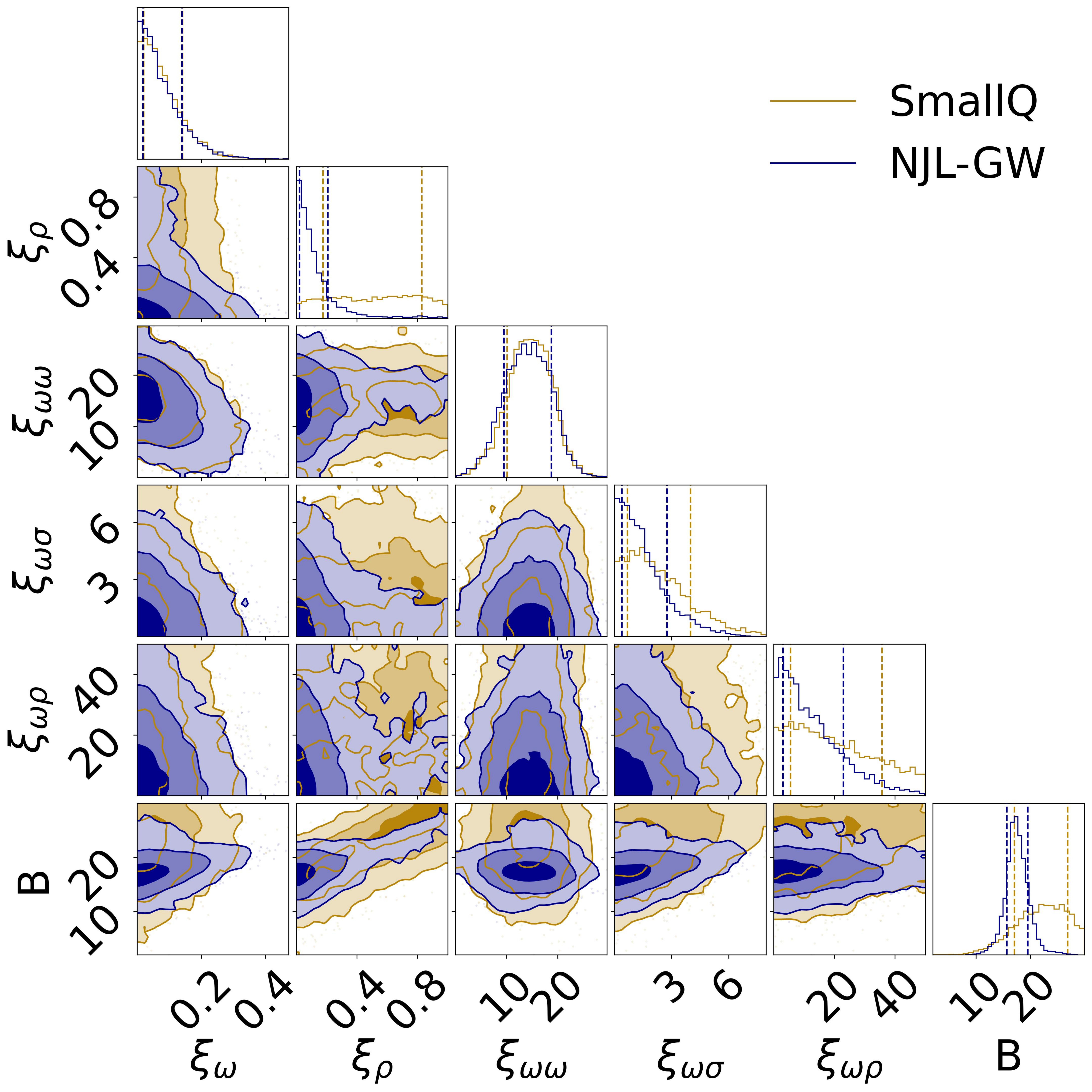}
      \end{tabular}
      \caption{Corner plot of hadron (left) and quark (right) parameters.}
      \label{fig:corner}
  \end{figure*}

    Figure \ref{fig:corner} shows the corner plot
    of the hadron (left plot) and quark parameters (right plot),
    for the NJL-GW (purple) and SmallQ (yellow) sets.
    Relaxing the constraints related to the size of the quark core at $\NSmax$
    affects all the hadron parameters, especially the $g_\sigma$ and $g_\omega$.
    These parameters are related to the $R_{1.4}$, as discussed in Section \ref{sec:results_ns}.
    The quark parameters are less affected, compared to the hadron parameters.
    We see notable differences in the $\xi_\rho$ and $B$ parameters.
    The former is related to the onset of the quark strange,
    and the latter, to the phase transition density and $\Mmax$,
    also discussed in Section \ref{sec:results_ns}.
  
%  \begin{figure*}[ht]
%      \centering
%      \includegraphics[width=.9\linewidth]{Images/corner_par_Q_smallQ.pdf}
%      \caption{Corner plot of quark parameters.}
%      \label{fig:cornerQ}
%  \end{figure*}

\bibliography{refs}

\end{document}